\documentclass[prd,superscriptaddress,nofootinbib,secnumarabic,preprintnumbers]{revtex4-2}[10pt]
\pdfoutput=1
\usepackage{latexsym}
\usepackage{mathrsfs, amsmath, amssymb}
\usepackage{diagbox, slashed, makecell}
\usepackage{hyperref, url}
\usepackage{color}
\usepackage{graphicx}
\usepackage{appendix}
\usepackage{verbatim}
\usepackage{multirow}
\usepackage{ucs}
\usepackage{indentfirst, layout, geometry}
\usepackage{enumitem}

\newcommand{\mr}{\mathrm}
\newcommand{\mc}{\mathcal}

\graphicspath{{figures/}}

\begin{document}

\begin{sloppypar}

\preprint{APCTP-Pre2024-017}

\title{Hybrid Type-II and Type-III seesaw model for the muon $g-2$ anomaly}

\author{Lei Cai}
\email{cailei@seu.edu.cn}
\affiliation{School of Physics, Southeast University, Nanjing 211189, P. R. China}

\author{Chengcheng Han}
\email{hanchch@mail.sysu.edu.cn}
\affiliation{School of Physics, Sun Yat-Sen University, Guangzhou 510275, P. R. China}
\affiliation{Asia Pacific Center for Theoretical Physics, Pohang 37673, Korea}

\author{Shi-Ping He}
\email{heshiping@tyut.edu.cn}
\affiliation{College of Physics and Optoelectronic Engineering, Taiyuan University of Technology, Taiyuan 030024, P. R. China}
\affiliation{Center for High Energy Physics, Peking University, Beijing 100871, P. R. China}
\affiliation{Asia Pacific Center for Theoretical Physics, Pohang 37673, Korea}

\author{Peiwen Wu}
\email{pwwu@seu.edu.cn}
\affiliation{School of Physics, Southeast University, Nanjing 211189, P. R. China}

\date{\today}

\begin{abstract}
In this work we investigate the muon anomalous magnetic dipole moment $a_\mu$ in a model that extends the Standard Model with a scalar triplet and a lepton triplet. Different from previous studies, we find that there is still viable parameter space in this model to explain the discrepancy $\Delta a_\mu=a_{\mu}(\mr{Exp})-a_{\mu}(\mr{SM})$.
While being consistent with the current data of neutrino mass, electroweak precision measurements and the perturbativity of couplings, our model can provide new physics contribution $a_\mu^\textrm{NP}$ to cover the central region of $\Delta a_\mu$ with new scalar and lepton mass as low as around TeV. This mass scale is allowed by the current collider searches for doubly charged scalars and the lepton triplet, and they can be tested at future high energy and/or high luminosity colliders.
\end{abstract}

\maketitle

\tableofcontents
\clearpage

\section{Introduction}

The anomalous magnetic moment (AMM) of muon, denoted as $a_{\mu}\equiv(g-2)_\mu/2$, has been theoretically predicted in the Standard Model (SM) and experimentally measured both to very high precision. Since the $g-2$ experiment E821 performed at Brookhaven National Laboratory
(BNL) released its data about two decades ago \cite{Muong-2:2006rrc}, a 
discrepancy $\Delta a_\mu=a_{\mu}(\mr{Exp})-a_{\mu}(\mr{SM})$ has been existing and triggered rich phenomenological studies (see reviews \cite{Jegerlehner:2009ry,Jegerlehner:2017gek} and references therein). 
Recently, on the theoretical side, a comprehensive summary of the most accurate SM prediction for $a_\mu$ is provided in \cite{Aoyama:2020ynm} where the value is reported as $a_{\mu}(\mr{SM})=116591810(43)\times10^{-11}$. 
On the experimental side, the Muon $g-2$ Experiment at Fermilab released its Run-1 dataset in 2021 and the result was $a_{\mu}^{2021}(\mr{Exp}) = 116592061(41)\times10^{-11}$ \cite{Muong-2:2021ojo} when combining the data from Fermilab Run-1 and BNL. This resulted in a deviation from the SM prediction of $a_{\mu}^{2021}(\mr{Exp})-a_{\mu}(\mr{SM}) = (251\pm59)\times10^{-11}$ with a significance of $4.2\, \sigma$.
In 2023, Fermilab released data from Run-2 and Run-3 leading to a new result after being combined with Run-1 and the BNL data as $a_{\mu}^{2023}(\mr{Exp}) = 116592059(22)\times10^{-11}$ \cite{Muong-2:2023cdq}. In this case the deviation from the SM prediction $\Delta a_\mu=a_{\mu}^{2023}(\mr{Exp})-a_{\mu}(\mr{SM}) = (249\pm48)\times10^{-11}$ achieved a significance of $5.1\,\sigma$ \footnote{Recently, there have been some discussions on the discrepancies in the calculation of the hadronic vacuum polarization contribution to $a_{\mu}(\mr{SM})$ between the lattice QCD calculations and experimental data in measurements of $e^- e^+ \to \pi^- \pi^+$, which seems to reduce  $\Delta a_\mu$ to some extend \cite{Venanzoni:2023mbe,Kuberski:2024bcj, Boccaletti:2024guq}. }. 
\footnote{During review of this paper, the authors update the SM prediction of $a_{\mu}$ \cite{Aliberti:2025beg}. Here, we will still adopt the value in paper \cite{Aoyama:2020ynm}.}
These deviations inspired phenomenological investigations in various new physics models. Typical examples include the two-Higgs-doublet model \cite{Cherchiglia:2017uwv, Iguro:2023tbk}, the dark photon model \cite{Pospelov:2008zw}, the supersymmetric models \cite{Moroi:1995yh, Stockinger:2006zn,Cox:2021nbo,Cox:2018vsv,Han:2020exx,Han:2021ify, Stockinger:2006zn}, leptoquark models \cite{He:2021yck, He:2022zjz, He:2022glw}, and vector-like lepton extended models \cite{Crivellin:2018qmi, Crivellin:2021rbq}. More references can be found in reviews \cite{Jegerlehner:2009ry,Jegerlehner:2017gek,Freitas:2014pua, Kowalska:2017iqv, Calibbi:2018rzv,Athron:2021iuf}.

The chiral structure of the $a_\mu$ can be described by the following effective tensor operator
\begin{align}\label{eq-Leff-mug2}
& \mathcal{L}^{\rm AMM}_{\rm eff.} = - \frac{e_\ell a_\ell}{4 m_\ell} \, \overline{\ell_L}\, \sigma^{\mu \nu} \, \ell_R \, F_{\mu \nu}\, + {\rm h.c.} \, ,
\end{align}
in which the SM charged lepton $\ell$ should be understood as muon flavor. We can see that both the left-handed (LH) $\ell_L$ and the right-handed (RH) $\ell_R$ of muon are involved. If one considers the new physics contribution to $a_\mu$ at 1-loop level matching the required chiral structure, proper chiral flip is needed along the fermion line. This chiral flip can take place either in the external muon line realized by the SM Yukawa interaction outside the loop or in the internal fermion line inside the loop \cite{Calibbi:2018rzv}.
In the framework of simplified models, contributions to $\Delta a_\mu$ can be generated at the 1-loop level by introducing several new physics fields with various spins and quantum numbers under the electroweak gauge group ${\rm SU(2)_L \times U(1)_Y}$. 

The new physics models trying to interpret  $\Delta a_\mu$ can also generate corrections to the muon mass, rare decays in the leptonic flavor sector and other electroweak precision observables. As pointed out in recent reviews \cite{Freitas:2014pua, Kowalska:2017iqv, Calibbi:2018rzv,Athron:2021iuf} many simplified models have been reported to fail to explain $\Delta a_\mu$ either due to the wrong sign of predicted $\Delta a_\mu$ or excluded by certain experimental constraints. Given the fact that the LH charged lepton and the LH neutral neutrino are embedded in the same ${\rm SU(2)_L}$ doublet, the possible interplay of generating the neutrino mass and explaining $\Delta a_\mu$ have intrigued rich discussions. Taking the Type-I \cite{Minkowski:1977sc, Yanagida:1979as, Gell-Mann:1979vob, Mohapatra:1979ia, Glashow:1979nm}, Type-II \cite{Magg:1980ut,Schechter:1980gr,Wetterich:1981bx,Lazarides:1980nt,Mohapatra:1980yp,Cheng:1980qt}, Type-III \cite{Foot:1988aq,Ma:2002pf} seesaw as examples, it has been well known that each individual of them generates ${\textit{negative}}$ contribution to $a_\mu$ and thus cannot explain $\Delta a_\mu > 0$ (details can be found in e.g. \cite{Chao:2008iw,Freitas:2014pua,Lindner:2016bgg} for Type-I, \cite{Freitas:2014pua,Lindner:2016bgg} for Type-II and \cite{Biggio:2008in,Lindner:2016bgg} for Type-III).

Given the failure of the simplest seesaw models to explain $\Delta a_\mu$, one can consider whether the hybrid seesaw models could accommodate the $\Delta a_\mu$. In this work we focus on the hybrid Type-II and Type-III seesaw model and we find that this model could explain the $\Delta a_\mu$ within 2$\sigma$. We note that our result is opposite to the conclusion of existing literature \cite{Freitas:2014pua, Calibbi:2018rzv} for the same particle contents. The reason causing the difference is that the following interactions are included in our analysis, while not included in \cite{Freitas:2014pua}\footnote{The reference \cite{Calibbi:2018rzv} forbids these two operators by imposing a $Z_2$ symmetry.}.
\begin{align}\label{eqn-operator-mass-mixing}
    \overline{L_L} S \epsilon(L_L)^C \quad , \quad \overline{L_L}(F_L)^{C}\epsilon H^{\ast} \, ,
\end{align}
where $S$ denotes the scalar triplet in Type-II seesaw (same as $\phi_T$ in \cite{Freitas:2014pua}) and $F_L$ denotes the fermion triplet in Type-III seesaw (same as $\psi_A$ in \cite{Freitas:2014pua}).
$S$ takes a quantum number $(3,\,-1)$ and $F_L$ takes a quantum number $(3,\, 0)$  under the SM electroweak gauge group ${\rm \mathrm{SU(2)_L}\times \mathrm{U(1)_Y}}$. $H, \, L_L$ are the SM Higgs doublet and LH lepton doublet, while $\epsilon\equiv i\sigma^2$ is the antisymmetric tensor. $(F_L)^C$ denotes the charge conjugation of $F_L$ satisfying $(F_L)^C\equiv C\overline{F_L}^T$. Note that the above two operators are dimension-4 and they will generate the mixing between fermion triplet $F_L$ and SM leptons. More importantly, these operators provide new source of chiral flip in the one-loop Feynman diagram of $a_\mu$. Since this new chiral flip takes place on the heavy fermion line (see the middle panel of Fig. \ref{fig-muongm2VLL_Feynman}), it will produce chiral enhancement $\propto M_{F_L}$ to $a_\mu$ which is helpful for our model to explain $\Delta a_\mu$.

This paper is organized as follows. In Sec. \ref{sec-model-setup-relevant-physics} we provide our model setup, an overview of the main physics and the dominant constraints on the parameter space.
In Sec. \ref{sec-analytical-numerical-results} we present our analytical and numerical results. We draw our conclusion in Sec.~\ref{sec-conclusion}. More details of our calculation and parameter choices can be found in the Appendices.

\section{Model setup and relevant physics }\label{sec-model-setup-relevant-physics}

\subsection{Model setup}\label{subsec-model-setup}
In our model we extend the SM with a scalar triplet $S$ and a left-handed lepton triplet $F_L$, of which the representations $(n,\, Y)$ under the SM electroweak gauge group ${\rm \mathrm{SU(2)_L}\times \mathrm{U(1)_Y}}$ are
\begin{align}
(n_S, \, Y_S)&=(3,\,-1) \, , \qquad (n_{F_L}, \, Y_{F_L})=(3,\, 0) \, ,
\end{align}
and the component fields are
\begin{align}
S\equiv\Big[\begin{array}{cc}\delta^-/\sqrt{2} &\quad  (v_{\delta}+\delta^0+i\, a^0)/\sqrt{2}\\ \delta^{--}&-\delta^-/\sqrt{2}\end{array}\Big] \, ,\qquad F_L \equiv\Big[\begin{array}{cc}F_{L}^0/\sqrt{2}&F_{L}^+\\ F_{L}^-&-F_{L}^0/\sqrt{2}\end{array}\Big] \, ,
\end{align}
in which $v_\delta$ is the vacuum expectation value (vev) of the neutral component of $S$ after electroweak symmetry breaking (EWSB).
We consider the following mass and Yukawa terms in the Lagrangian which are most relevant to the physics of $(g-2)_\mu$
\begin{equation}
\begin{aligned}
\mc{L}_\textrm{mass+Yuk.}\supset & -\frac{1}{2}M_F\mr{Tr}\left[\overline{F_L}(F_L)^{C}\right]-y^{ij}\overline{L_L^i}\ell_R^jH \\
& -x_L^{ij}\overline{L_L^i}S\epsilon(L_L^j)^C-\lambda_L^i\overline{\ell_R^i}\mr{Tr}\left[F_L S\right]-z_L^i\overline{L_L^i}(F_L)^{C}\epsilon H^{\ast}+\mathrm{h.c.} \, ,
\end{aligned}\label{eq-LYukawa-FL}
\end{equation}
in which $H$ is the SM Higgs doublet and $\epsilon\equiv i\sigma^2$ is the antisymmetric tensor. $L^{i/j}_L\equiv(\nu^{i/j}_L,\, \ell^{i/j}_L)^T$  and $\ell^{i/j}_R$ are the LH doublet and RH singlet of SM lepton under ${\rm \mathrm{SU(2)_L}}$ with $i,j=1,2,3$ denoting the generation index, respectively.
$(F_L)^C$ denotes the charge conjugation of $F_L$ satisfying $(F_L)^C\equiv C\overline{F_L}^T$. We require all Yukawa couplings in our model to be real in order to avoid constraints from CP-violating observables.

After EWSB we have the expansion $H=[G^+,({v_h}+h+i\, G^0)/\sqrt{2}]^T$ and can derive the following neutrino and charged lepton mass matrices
\begin{equation}
\begin{aligned}\label{eq-lepton-mass-mixing-neutral}
\mathcal{L}_\textrm{mass}= &-\frac{1}{2}
\left[\begin{array}{cc} \overline{\nu_L^i}&\overline{F_L^0} \end{array}\right]
\left[\begin{array}{cc}-\sqrt{2} x_L^{ij}v_{\delta}&\frac{1}{2}z_L^i{v_h} \\ \frac{1}{2}z_L^i{v_h} & M_F \end{array}\right]
\left[\begin{array}{cc}(\nu_L^j)^C\\(F_L^0)^C \end{array}\right]\\
&-
\left[\begin{array}{cc} \overline{\ell_L^i}&\overline{F_L^-} \end{array}\right]
\left[\begin{array}{cc}\frac{1}{\sqrt{2}}y^{ij}{v_h}&\frac{1}{\sqrt{2}}z_L^i{v_h} \\ \frac{1}{\sqrt{2}}(\lambda_L^i)^{\ast}v_{\delta} & M_F \end{array}\right]
\left[\begin{array}{cc}\ell_R^j\\(F_L^+)^C \end{array}\right]+\mathrm{h.c.} \, .
\end{aligned}
\end{equation}
For simplicity, we consider the scenario that only the second generation $i,j\equiv 2$ exists in Eq. \eqref{eq-LYukawa-FL} to generate the mass mixing between muon flavor and heavy leptons in the new physics sector. Therefore, the indices $i,j$ will be dropped in the following. We can diagonalize the previous mass matrices through the following rotations
\begin{equation}
\begin{aligned}\label{eqn:model:1S1VLL:FL:rotation}
\left[\begin{array}{c}\nu_L\\F_L^0\end{array}\right] &\rightarrow
	\left[\begin{array}{cc}c_L^{\nu}&s_L^{\nu}\\-s_L^{\nu}&c_L^{\nu}\end{array}\right]
	\left[\begin{array}{c}\nu_L\\F_L^0\end{array}\right] \, , \\
\left[\begin{array}{c}\ell_L\\F_L^-\end{array}\right] &\rightarrow
	\left[\begin{array}{cc} c_L^\ell & s_L^\ell \\- s_L^\ell & c_L^\ell \end{array}\right]
	\left[\begin{array}{c}\ell_L\\F_L^-\end{array}\right] \, , \\
	\left[\begin{array}{c}\ell_R\\(F_L^+)^C\end{array}\right] &\rightarrow
	\left[\begin{array}{cc} c_R^\ell & s_R^\ell \\- s_R^\ell & c_R^\ell \end{array}\right]
	\left[\begin{array}{c}\ell_R\\(F_L^+)^C\end{array}\right] \, .
\end{aligned}
\end{equation}
In the above, $s_{L/R}^{\nu/\ell}$ and $c_{L/R}^{\nu/\ell}$ are abbreviations of $\sin\theta_{L/R}^{\nu/\ell}$ and $\cos\theta_{L/R}^{\nu/\ell}$ when applicable. In Appendix \ref{appendix-sec-Details-of-Lagrangian-terms-and-interactions}, we present details of the Yukawa interactions in mass eigenstates. 

As for the scalar sector, there can be rich interactions involving $H$ and $S$ \cite{Arhrib:2011uy} as follows
\begin{align}\label{eq-Lag-VHS}
V(H,S)=&-m_H^2H^{\dag}H+\frac{\lambda}{4}(H^{\dag}H)^2+M_S^2 \mathrm{Tr}[S^{\dag}S]+[\mu_{HS}(H^T \epsilon SH)+\mathrm{h.c.}]\nonumber\\
& +\lambda_1(H^{\dag}H)\mathrm{Tr}[S^{\dag}S]+\lambda_2 (\mathrm{Tr}[S^{\dag}S])^2+\lambda_3 \mathrm{Tr}[(S^{\dag}S)^2]+\lambda_4 H^{\dag}S^{\dag}S H   \, ,
\end{align}
which would generate the mixing between $H$ and $S$ and result in the mass eigenstates including the SM Higgs boson and several scalars in the new physics sector. Similar to Eq. \eqref{eqn:model:1S1VLL:FL:rotation} we can diagonalize the electrically neutral CP-even scalar mass matrix by performing the following rotations
\begin{equation}\label{eq_scalar_rotation}
\begin{aligned}
\left[\begin{array}{c}h\\ \delta^0\end{array}\right]&\rightarrow
	\left[\begin{array}{cc}c^h &s^h\\-s^h&c^h\end{array}\right]
	\left[\begin{array}{c}h\\\delta^0\end{array}\right] \, , \\
\left[\begin{array}{c}G^0 \\ a^0 \end{array}\right]&\rightarrow
	\left[\begin{array}{cc}c^a &s^a\\-s^a&c^a\end{array}\right]
	\left[\begin{array}{c}G^0\\ a^0\end{array}\right] \, , \\
 \left[\begin{array}{c} (G^+)^\ast \\ \delta^- \end{array}\right]&\rightarrow
	\left[\begin{array}{cc}c^G &s^G\\-s^G&c^G\end{array}\right]
	\left[\begin{array}{c}(G^+)^\ast  \\ \delta^- \end{array}\right] \, ,
\end{aligned}
\end{equation}
in which $s^{h/a/G}, c^{h/a/G}$ are the abbreviations of $\sin\theta^{h/a/G}, \cos\theta^{h/a/G}$ following the similar convention as Eq. \eqref{eqn:model:1S1VLL:FL:rotation}, and $G^0,G^\pm$ are the Goldstone bosons which will be absorbed to be the longitudinal components of SM gauge bosons $Z,W^\pm$ via the Higgs mechanism.
For later analysis we will label the masses of physical scalars in the mass eigenstates as $m_h, m_{\delta^0}, m_{a^0},m_{\delta^-},m_{\delta^{--}}$, with $h,\delta^0$ being the neutral CP-even scalars, $a^0$ being the neutral CP-odd scalar, and $\delta^-,\delta^{--}$ being the singly and doubly charged scalars, respectively. Details of the diagonalization and rotation matrices in the scalar sector can be found in e.g. \cite{Arhrib:2011uy}. In Appendix \ref{appendix_subsec_Input_parameters}, we show the discussions on the scalar sector parameters.

\begin{figure}
    \centering
    \includegraphics[width= 0.8 \linewidth]{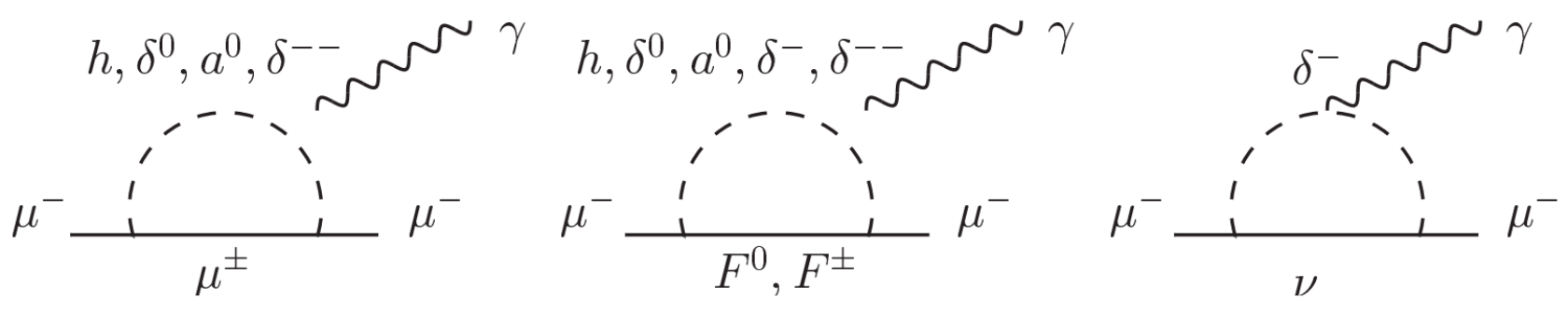}
    \caption{Main Feynman diagrams contributing to the $(g-2)_{\mu}$ in our model beyond the SM, in which the left, middle and right panel corresponds to $a_{\mu}^{\ell,\,\textrm{total}}, \, a_{\mu}^{F,\,\textrm{total}} , \, a_{\mu}^{\nu,\,\textrm{total}}$ in Eq. \eqref{eq-aNP}, respectively. In the left and middle diagrams the photon can be emitted from either of the charged particles in the loop, while the emission only takes place from $\delta^-$ in the right diagram. Our calculations show that the middle panel with $\delta^-, \delta^{--}$ are dominant contributions. 
    See Appendix \ref{appendix-sec-analytica-results} for details. The Feynman diagrams are drawn using \textbf{JaxoDraw} \cite{Binosi:2008ig}.}
    \label{fig-muongm2VLL_Feynman}
\end{figure}

\subsection{Main physical implications}\label{subsec_overview_main_physics}
In this subsection we provide an overview of the main physics of our model.

\begin{itemize}
\item {\textbf{New physics contribution to $a_\mu$}}

The new fields in our model provide new contributions to $a_\mu$ which we will denote as $a^{\rm NP}_\mu$ hereafter. Fig. \ref{fig-muongm2VLL_Feynman} shows the main 1-loop processes generating $a^{\rm NP}_\mu$ in terms of mass eigenstates originating from the new fields. Our calculations show that the middle panel with $\delta^-, \delta^{--}$ running in the loop are dominant contributions compared to the remaining diagrams in Fig. \ref{fig-muongm2VLL_Feynman}.
Note that internal scalars in Fig. \ref{fig-muongm2VLL_Feynman} can be replaced by SM gauge bosons when applicable, However, the corresponding contributions are negligible compared to the scalar cases. See Appendix \ref{appendix-sec-analytica-results} for details.

\item {\textbf{Neutrino mass generation from hybrid seesaw}}

The neutrino mass in our hybrid seesaw model is determined after diagonalizing the neutral lepton mass matrix in Eq. \eqref{eq-lepton-mass-mixing-neutral} as follows
\begin{align}
\label{eq-neutrino-mass-analytical-expression}
m_\nu = \frac{1}{2} [(M_F - \sqrt{2} x_L v_\delta) - \sqrt{(M_F + \sqrt{2} x_L v_\delta)^2 + z_L^2 v_h^2} ] \approx - \sqrt{2} x_L v_\delta - \frac{z_L^2 v_h^2}{4 M_F},
\end{align}
in which $v_\delta \approx \mu_{HS} v_h^2 /\sqrt{2} M_S^2$ after EWSB. This implies that neutrino masses are generated  by both Type-II (first term in the R.H.S of Eq.\,\eqref{eq-neutrino-mass-analytical-expression}) and Type-III seesaw mechanism (second term in R.H.S of Eq.\,\eqref{eq-neutrino-mass-analytical-expression}). 
Since the neutrino mass is order of 0.1 eV, we require some fine-tuning to get the correct neutrino masses in our parameter space. We can define the following quantity to measure the degree of fine-tuning:
\begin{align}
\Delta_{FT}\equiv \frac{m_{\nu}}{x_Lv_{\delta}},
\end{align}
which turns out to be $\Delta_{FT}\sim\mc{O}(10^{-9})$ when considering neutrino mass being at order of 0.1 eV and the parameter space in our numerical analyses. Such fine-tuning is difficult to justify by invoking additional symmetries. However, the fine-tuning of the neutrino mass might be explained by the anthropic principle~\cite{Tegmark:2003ug}, as a neutrino mass larger than $\mathcal{O}(1)$ eV would destroy the large-scale structure of our universe.

\item {\textbf{Correction to muon mass}}

Diagrams in Fig. \ref{fig-muongm2VLL_Feynman} without emission of photon will generate corrections to muon mass. In the parameter space favored by explaining $\Delta a_\mu$, we find that the mass corrections in our model is $\delta m/m_{\mu}\sim\mc{O}(1)$. See Appendix \ref{sec-estimation-corrections-to-muon-mass} for details.
\end{itemize}

\subsection{Choices of input parameters and main constraints}\label{subsec-constraints-on-parameter-space}

In this subsection, to highlight the key relations among model parameters utilized in presenting our main results, while keeping the readability of the manuscript, we quickly go through our setup of choosing independent input parameters and the main constraints on the parameter space of our model. We 
mainly discuss the fermion sector here since it contains the chiral enhancement effects which are essential to our main goal of explaining $\Delta a_\mu$. 
More detailed are given in Appendix \ref{appendix_subsec_Relation_of_parameters} and Appendix \ref{appendix_subsec_Input_parameters}.

In the fermion sector shown in Eq. \eqref{eq-LYukawa-FL}, we can convert the five parameters in the Lagrangian
\begin{align}
M_F \, , \quad y \, , \quad x_L \, , \quad \lambda_L \, , \quad z_L \, ,
\end{align}
to physical quantities in terms of mass eigenstates as follows
\begin{align}
m_{\nu}, \, m_{F^0}, \, s^\nu_L &: \quad \textrm{from first line of Eq. \eqref{eq-lepton-mass-mixing-neutral}} \, ,  \nonumber \\
m_{\ell}, \, m_{F^\pm}, \, s^\ell_L, \, s^\ell_R &: \quad \textrm{from second line of Eq. \eqref{eq-lepton-mass-mixing-neutral}}\, .
\end{align}
Despite there are seven quantities listed above, two of them are not independent as shown in Eq. \eqref{eq-remove-two-extra-dof}.
As for the five independent parameters, three of them can be naturally chosen as
\begin{align}
m_{\nu} \, , \quad \, m_{\ell} \, , \, \quad m_{F^0}  \, .
\end{align}
According to Eq. \eqref{eq-small-mixing-angle-RH} and similar to Eq. \eqref{eq-notation-mass-delta}, hereafter we would use $m_F$ as the following notation 
    \begin{align}\label{eq-notation-mass-F}
    m_F \equiv m_{F^0} \approx m_{F^\pm} \, ,
    \end{align}
to denote the numerical approximations, while keeping in mind they can be independent quantities in principle as discussed above. As discussed in Eq. \eqref{eq-mFpm-mF0-lower-bound}, in this work we would take
    \begin{align}\label{eq-mF-lower-bound}
        m_F \gtrsim 1000 \, {\rm GeV} \, .
    \end{align}   
    
To better illustrate our results and relevant physical features, in the following we consider the following two different schemes of independent input parameters.
\begin{itemize}
\item {$\{\lambda_L, x_L\}$  scheme}\\ \\
This scheme is physically suitable for the illustration of the decoupling behavior in the predictions of $(g-2)_\mu$ in our model. With more details provided in Appendix \ref{appendix_subsec_Input_parameters}, Eqs. \eqref{eqn:SFL:input:relationsapp-neutral} and \eqref{eqn:SFL:input:relationsapp-charged} suggest that
the other relevant parameters can be expressed as
\begin{align}\label{eq-xxL-as-input-Yukawa-expression}
& z_L = 2\sqrt{ - \sqrt{2} \, x_L \frac{v_\delta \, m_F}{v_h^2} } \, , \quad  y \approx  2\, \lambda_L \frac{v_\delta}{v_h} \sqrt{\frac{- x_L \, v_\delta}{\sqrt{2} \, m_F}} \, , \nonumber \\
& s^\nu_L \approx \sqrt{ - \sqrt{2} \, x_L  \frac{v_\delta}{m_F} } \, , \qquad \, s^\ell_L \approx \sqrt{2} \, s^\nu_L \, , \qquad  s^\ell_R \approx \frac{\lambda_L \, v_\delta}{\sqrt{2} \, m_F} \, .
\end{align}
Note that in our numerical analysis we choose all Yukawa couplings to be real and in our conventions we have $x_L < 0$ and $y, z_L, \lambda_L> 0$.
\item {$\{\lambda_L, s_L^{\nu}\}$ scheme} \\ \\
This scheme aims to highlight the chiral enhancement effects originating from the existence of heavy new fermions in the loop generating $(g-2)_\mu$. While more details are given in Appendix \ref{appendix_subsec_Input_parameters}, the other relevant parameters can be expressed as the following via Eqs. \eqref{eqn:SFL:input:relationsapp-neutral} and \eqref{eqn:SFL:input:relationsapp-charged},
\begin{align}\label{eq-snL-as-input-Yukawa-expression}
& x_L \approx - \frac{1}{\sqrt{2}} (s^\nu_L)^2\frac{m_F}{v_\delta} \, , \quad  z_L \approx 2 \, s^\nu_L \frac{m_F}{v_h} \, , \quad y\approx \sqrt{2} \, s^\nu_L  \, \lambda_L \frac{v_\delta}{v_h} \, ,\nonumber \\
&s^\ell_L \approx \sqrt{2} \, s^\nu_L \, , \quad\quad\quad\quad\quad  s^\ell_R  \approx \frac{\lambda_L \, v_\delta}{\sqrt{2} \, m_F} \, .
\end{align}
\end{itemize}

Combining the above discussion with other details in Appendix \ref{appendix_subsec_Relation_of_parameters} and Appendix \ref{appendix_subsec_Input_parameters}, the input parameters in our analysis are summarized as follows
\begin{align}
\textbf{\textrm{Fixed}} : &\quad m_{\nu}=0 \, , \quad\quad\,\,\,\,   m_\mu =105.66 \, \mathrm{MeV} \, , \quad   m_h =125 \, \mathrm{GeV} \, ,  \quad v_h \approx 246 \, \textrm{GeV} \, , \nonumber\\
&\quad v_{\delta} = 5\, \textrm{GeV} \, ,  \quad m_{\delta}= 1000 \, \textrm{GeV} \, , \quad\quad s^h=s^a=s^G=0 \, ,  \nonumber\\
\textbf{\textrm{Varying}} : &\quad    m_F \, \qquad\qquad\,\,\, \textrm{ and } \qquad\qquad\qquad\quad\, \{\, \lambda_L,\,  x_L \, \} \, \textrm{ or } \, \{\,  \lambda_L, \, s_L^{\nu} \, \} \, .
\end{align}

Meanwhile, the main constraints on the parameter space of our model include the following ones.
\begin{itemize}
\item {\textbf{Limits on the Lepton Flavor Violating (LFV) rare decays}}
    
    As mentioned below Eq. \eqref{eq-lepton-mass-mixing-neutral}, we consider the scenario that only the second generation $i, j = 2$ in $x^{ij}_L, z_L^i, \lambda^i_L$ are nonzero, which will make our model safe in light of the various LFV constraints.    
\item {\textbf{Limits on the mixing between SM and heavy neutrinos }}
    
    $s^\nu_L$ characterizing the mixing of SM neutrino $\nu$ with $F^0$ is constrained by the electroweak precision measurements, for the muon flavor to be \cite{delAguila:2008pw},
\begin{align}\label{eq-snL-exp-constraint}
s^\nu_L \leq 0.017 .
\end{align} 
\item {\textbf{Limits from precision measurements of SM Higgs}}

        As for the mixing $s^h,s^a,s^G$ of the scalar sector in Eq. \eqref{eq_scalar_rotation}, we first have the following requirement to safely pass the current constraints from the study of Higgs data \cite{ATLAS:2021vrm,CMS:2020gsy}
    \begin{align}\label{eq-sh-value}
        |s^h| \lesssim \mathcal{O}(0.1).
    \end{align}
    We will simply impose the following simplifications in numerical calculation which make negligible difference to the results 
        \begin{align}\label{eq-sh-value}
            s^h=s^a=s^G=0 \, .
        \end{align}
     See Appendix \ref{appendix_subsec_Input_parameters} for details.
\item {\textbf{Requirement of the perturbativity of Yukawa couplings}}
    
    We require all Yukawa couplings $x_L, z_L, y, \lambda_L$ in Eq. \eqref{eq-LYukawa-FL} to be less than $\mathcal{O}(1)$. In our numerical analysis, we choose the value of input parameters satisfying the requirement of perturbativity. See Appendix \ref{appendix_subsec_Perturbativity_requirement} for details.
\end{itemize}

\section{Analytical and numerical results}\label{sec-analytical-numerical-results}

In this section we present our main results. We cross check our formulae by implementing our model in Eq. \eqref{eq-LYukawa-FL} to \textbf{FeynRules} \cite{Christensen:2008py, Alloul:2013bka} interfaced to \textbf{FeynArts} \cite{Hahn:2000kx} and \textbf{FormCalc} \cite{Hahn:2016ebn} to perform loop calculations. Then we extract from the amplitude to obtain the expressions of $a^\textrm{NP}_\mu$, i.e. the new physics contribution to $\Delta a_\mu$ in our model, and further reduce the loop functions to simple expressions via \textbf{Package-X} \cite{Patel:2015tea, Patel:2016fam}.

\subsection{Analytical results}\label{subsection_Analytical_results}

In this subsection we highlight our main analytical results while leaving more details of calculation in Appendix \ref{appendix-sec-analytica-results}.

In Fig. \ref{fig-muongm2VLL_Feynman} we show the Feynman diagrams contributing to $(g-2)_{\mu}$ in our model, in which the leptons $\ell, \nu$ should be understood to carry the muon flavor as $\mu, \nu_\mu$. They are divided into three parts according to the fermions appearing in the loop. Originating from the left, middle and right panel, respectively, the analytical results of $a^\textrm{NP}$ generated by our model can be decomposed to
\begin{align}\label{eq-aNP}
a^\textrm{NP}_{\mu}\equiv a_{\mu}^{\ell,\,\textrm{total}} +a_{\mu}^{F,\,\textrm{total}} +a_{\mu}^{\nu,\,\textrm{total}} \, .
\end{align}
Note that all Yukawa couplings in our model are chosen to be real, but we will present our analytical results in the context of complex Yukawa couplings for the more general scenario. To write down the analytical results, we first define the following expressions as the reduced form of loop functions (see \cite{He:2023fkv} for a more complete list).
\begin{itemize}
\item In calculating $a_{\mu}^{\ell,\,\textrm{total}}$ from the left panel of Fig. \ref{fig-muongm2VLL_Feynman} which satisfies $m_\ell \ll m_h,\, m_\delta $, we define
\begin{align}
F_{LL}^{f,1}(x)&=\frac{1}{6}+x(\frac{1}{2}\log x+\frac{25}{24})\, ,\quad F_{LR}^{f,1}(x)=-\frac{1}{2}\log x-\frac{3}{4}+x(-2\log x-\frac{8}{3})\, ,\nonumber\\
F_{LL}^{S,1}(x)&=-\frac{1}{12}+\frac{1}{8}x \, ,\qquad\qquad\quad\, F_{LR}^{S,1}(x)=-\frac{1}{4}+x(-\frac{1}{2}\log x-\frac{11}{12}) \, .
\end{align}
\item In calculating $a_{\mu}^{F,\,\textrm{total}}$ from the middle panel of Fig. \ref{fig-muongm2VLL_Feynman} which satisfies $m_\ell \ll m_F,\, m_\delta$, we define
\begin{align}\label{eq-loop-function-class-2}
F_{LL}^{f,2}(x)&=\frac{2+3x-6x^2+x^3+6x\log x}{12(1-x)^4} \, ,\quad\quad\,\,\,\, F_{LR}^{f,2}(x)=\frac{-3+4x-x^2-2\log x}{4(1-x)^3} \, ,\nonumber\\
F_{LL}^{S,2}(x)&=-\frac{1-6x+3x^2+2x^3-6x^2\log x}{12(1-x)^4}\, , \quad F_{LR}^{S,2}(x)=\frac{-1+x^2-2x\log x}{4(1-x)^3} \, .
\end{align}
\item In calculating $a_{\mu}^{\nu,\,\textrm{total}}$ from the right panel of Fig. \ref{fig-muongm2VLL_Feynman} which satisfies $m_\nu \ll m_\ell \ll m_\delta $, we define
\begin{align}
&F_{LL}^{S,3}(x)=-\frac{1}{12}(1+\frac{1}{2}x)\, ,\quad F_{LR}^{S,3}(x)=-\frac{1}{4}(1+\frac{2}{3}x) \, .
\end{align}
\end{itemize}

As a result of the chiral enhancement effects, $a_{\mu}^{F,\,\textrm{total}}$ in Eq. \eqref{eq-aNP} turns out to play the absolutely dominant role compared to $a_{\mu}^{\ell,\,\textrm{total}}$ and $a_{\mu}^{\nu,\,\textrm{total}}$, i.e. 
\begin{equation}\label{eq-ratio-inside-aeF}
    a_{\mu}^{F,\,\textrm{total}} \, \gg  \, a_{\mu}^{\ell,\,\textrm{total}} \, , \, a_{\mu}^{\nu,\,\textrm{total}} \, .
\end{equation}
Furthermore, $a_{\mu}^{F}$ originating from the middle panel of Fig. \ref{fig-muongm2VLL_Feynman} can be written as
\begin{align}\label{eq-mF-vs-xL-amueF}
a_{\mu}^{F,\,\textrm{total}}=a_{\mu}^{F,\,h}+a_{\mu}^{F,\,\delta^0}+a_{\mu}^{F,\,a^0}+a_{\mu}^{F,\,\delta^{--}}+a_{\mu}^{F,\,\delta^{-}} \, ,
\end{align}
in which we find that $a_{\mu}^{F,\,\delta^{--}}$ and $a_{\mu}^{F,\,\delta^{-}}$ contribute the dominant and sub-dominant part, respectively. Their expressions are
\begin{align}\label{eq-mF-vs-xL-amueF-charged}
a_{\mu}^{F,\,\delta^{--}}&=\quad\frac{m_{\mu}^2}{8\pi^2m_{\delta^{--}}^2}[4|x_L |^2 (s_L^\ell  c_L^\ell)^2 +|\lambda_L|^2 \big( ( c_R^\ell )^2-( s_R^\ell )^2 \big)^2]\cdot[-F_{LL}^{f,2}(\frac{m_{F^{\pm}}^2}{m_{\delta^{--}}^2})+2F_{LL}^{S,2}(\frac{m_{F^{\pm}}^2}{m_{\delta^{--}}^2})]\nonumber\\
&\quad +\frac{m_{\mu}m_{F^{\pm}}}{4\pi^2m_{\delta^{--}}^2} \textrm{Re}[ (2x_L s_L^\ell  c_L^\ell  ) \big( \lambda_L^{\ast} (( c_R^\ell )^2-( s_R^\ell )^2 ) \big)  ] \cdot[-F_{LR}^{f,2}(\frac{m_{F^{\pm}}^2}{m_{\delta^{--}}^2})+2F_{LR}^{S,2}(\frac{m_{F^{\pm}}^2}{m_{\delta^{--}}^2})] \, , \nonumber\\
a_{\mu}^{F,\,\delta^{-}}&= \quad \frac{m_{\mu}^2}{8\pi^2m_{\delta^-}^2}[2|x_L|^2( s_L^{\nu}  c_L^\ell c^G)^2+|\lambda_L|^2( c_L^{\nu}  c_R^\ell c^G )^2]\cdot F_{LL}^{S,2}(\frac{m_{F^0}^2}{m_{\delta^{-}}^2})\nonumber\\
&\quad +\frac{m_{\mu}m_{F^0}}{4\pi^2m_{\delta^-}^2} \textrm{Re} [(\sqrt{2} x_L  s_L^{\nu} c_L^\ell c^G )(  \lambda_L^{\ast}  c_L^{\nu} c_R^\ell c^G )] \cdot F_{LR}^{S,2}(\frac{m_{F^0}^2}{m_{\delta^-}^2}) \, .
\end{align}
To simplify our analytical results and highlight the chiral enhancement, we utilize the small mixing condition $s^\nu_L, s^\ell_L, s^\ell_R,s^h,s^a,s^G \ll 1$ discussed in  \ref{appendix_subsec_Input_parameters} and extract the most relevant terms in Eq. \eqref{eq-mF-vs-xL-amueF} as follows
\begin{align}\label{eq-aNP-most-simplified-step-1}
a^\textrm{NP}_{\mu} \approx &\, a_{\mu}^{F,\,\textrm{total}} \approx \, a_{\mu}^{F,\,\delta^{--}}+a_{\mu}^{F,\,\delta^{-}}  \, , \nonumber \\
\approx &\quad\quad
\frac{m_{\mu}m_{F^{\pm}}}{4\pi^2m_{\delta^{--}}^2} \textrm{Re}[ (2x_L s_L^\ell  c_L^\ell  ) \big( \lambda_L^{\ast} (( c_R^\ell )^2-( s_R^\ell )^2 ) \big)  ] \cdot[-F_{LR}^{f,2}(\frac{m_{F^{\pm}}^2}{m_{\delta^{--}}^2})+2F_{LR}^{S,2}(\frac{m_{F^{\pm}}^2}{m_{\delta^{--}}^2})]\nonumber\\
&+ \frac{m_{\mu}m_{F^0}}{4\pi^2m_{\delta^-}^2} \textrm{Re} [(\sqrt{2} x_L  s_L^{\nu} c_L^\ell c^G )(  \lambda_L^{\ast}  c_L^{\nu} c_R^\ell c^G )] \cdot F_{LR}^{S,2}(\frac{m_{F^0}^2}{m_{\delta^-}^2}) \, , \nonumber \\
\approx &\quad \quad\frac{m_{\mu}m_{F^{\pm}}}{4\pi^2m_{\delta^{--}}^2} \textrm{Re}[ (2x_L s_L^\ell  ) ( \lambda_L^{\ast} )  ] \cdot[-F_{LR}^{f,2}(\frac{m_{F^{\pm}}^2}{m_{\delta^{--}}^2})+2F_{LR}^{S,2}(\frac{m_{F^{\pm}}^2}{m_{\delta^{--}}^2})]\nonumber\\
&+ \frac{m_{\mu}m_{F^0}}{4\pi^2m_{\delta^-}^2} \textrm{Re} [(\sqrt{2} x_L  s_L^{\nu} )(  \lambda_L^{\ast} )] \cdot F_{LR}^{S,2}(\frac{m_{F^0}^2}{m_{\delta^-}^2})  \, . 
\end{align}
Utilizing $s^\ell_L\approx \sqrt{2} s^\nu_L$ from Eq. \eqref{eq-small-mixing-angle-LH}, as well as the mass relations $m_\delta \equiv m_{\delta^0}\approx m_{\delta^-}\approx m_{\delta^{--}}$ from Eq. \eqref{eq-notation-mass-delta} and $m_F\equiv m_{F^{\pm}}\approx m_{F^0}$ from Eq. \eqref{eq-notation-mass-F}, we can have the following simple expressions
\begin{align}\label{eq-aNP-most-simplified-step-3}
& a^\textrm{NP}_{\mu} \approx 
\frac{\sqrt{2}m_{\mu}m_F}{4\pi^2m_\delta^2} \textrm{Re} [ x_L  \lambda_L^{\ast} ] s_L^{\nu} \cdot [-2F_{LR}^{f,2}(\frac{m_F^2}{m_\delta^2})+5F_{LR}^{S,2}(\frac{m_F^2}{m_\delta^2}) ] \, .
\end{align}
Note that only two of the parameters $x_L, \lambda_L, s_L^{\nu}$ are independent.
We also checked that the modification to our results is negligible when further considering the scalar and lepton mass splittings in the calculations, which is less than $\mathcal{O}(10^{-2})$.

\subsection{Numerical results}
\begin{figure}
    \centering
    \includegraphics[width= 0.48 \linewidth]{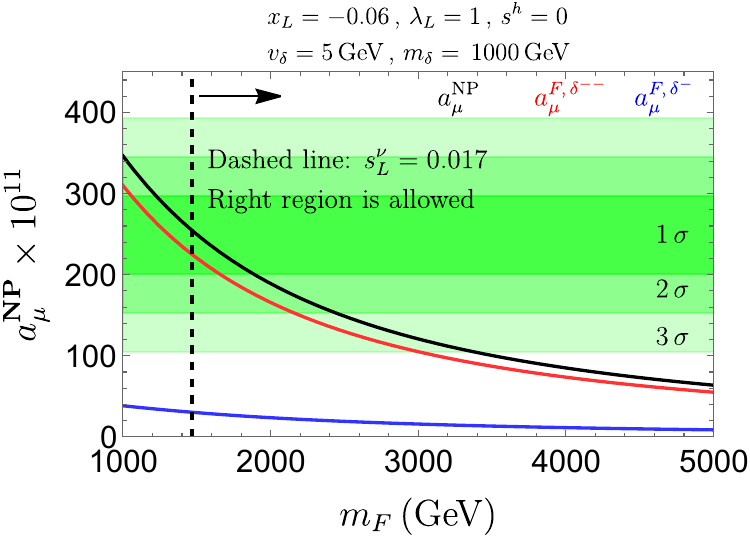}
        \includegraphics[width= 0.48 \linewidth]{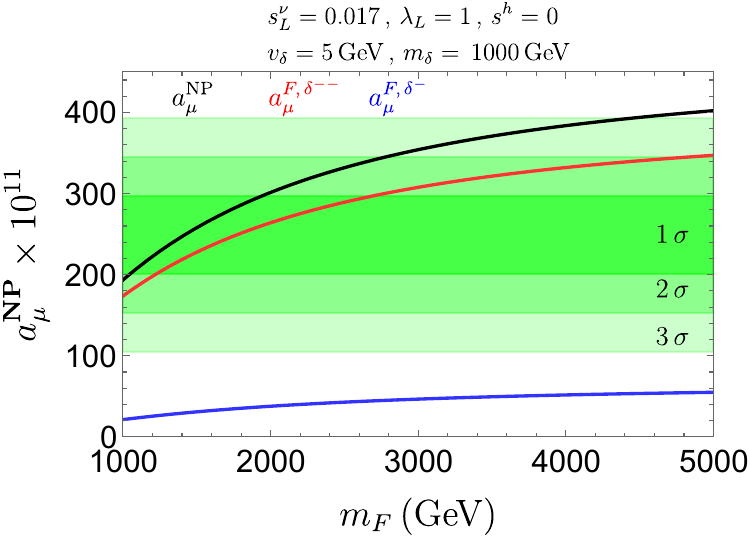}
    \caption{
    New physics contribution $a^\textrm{NP}_\mu$ (solid black line) predicted by our model. Dark, medium and light green color correspond to $1\sigma, \, 2\sigma, \, 3\sigma$ ranges of $\Delta a_\mu$. 
    Solid red (blue) line is $a_{\mu}^{F,\,\delta^{--}}$ ($a_{\mu}^{F,\,\delta^{-}}$) from Eq. \eqref{eq-mF-vs-xL-amueF} which is the dominant (sub-dominant) contribution to $a^\textrm{NP}_\mu$. 
    {\bf Left}: $\{\lambda_L, x_L\}$ as input parameters discussed in Eq. \eqref{eq-xxL-as-input-Yukawa-expression}. {\bf Right}: $\{\lambda_L, s^\nu_L\}$ as input parameters discussed in Eq. \eqref{eq-snL-as-input-Yukawa-expression}.}
    \label{fig-amuNP}
\end{figure}

In Fig.~\ref{fig-amuNP} we show the new physics contribution $a^\textrm{NP}_\mu$ predicted by our model in Eq. \eqref{eq-aNP} aiming at interpreting $\Delta a_\mu$ with green color of dark, medium and light opacity indicating the $1\sigma, \, 2\sigma, \, 3\sigma$ ranges of $\Delta a_\mu$. 
Solid black lines denote $a_{\mu}^{\textrm{NP}}$ in Eq. \eqref{eq-mF-vs-xL-amueF} which numerically satisfies $a_{\mu}^{\textrm{NP}} \approx a_{\mu}^{F,\,\textrm{total}}$ according to Eq. \eqref{eq-aNP-most-simplified-step-1}. Solid red and blue lines correspond to $a_{\mu}^{F,\,\delta^{--}}$ and $a_{\mu}^{F,\,\delta^{-}}$ in Eq. \eqref{eq-mF-vs-xL-amueF} which are the dominant and sub-dominant parts of $a_{\mu}^{F,\,\textrm{total}}$.

In the left panel of Fig. \ref{fig-amuNP} we impose the $\{\lambda_L, x_L \}$ scheme described in Eq. \eqref{eq-xxL-as-input-Yukawa-expression} as input parameters. 
We take $\lambda_L=1, \, x_L=-0.06$ as benchmark point and it can be seen that $a^\textrm{NP}_\mu$ predicted by our model can easily cover the central region of $\Delta a_\mu$ with $m_F\approx 1500 \, \textrm{GeV}$. In more details, $a_{\mu}^{F,\,\delta^{--}}$ ($a_{\mu}^{F,\,\delta^{-}}$) indicated by red (blue) line contribute about $85\%$ ($10\%$) to the total $a^\textrm{NP}_\mu$ in the mass range of $m_F$ shown in Fig. \ref{fig-amuNP}. 
Note that according to Eq. \eqref{eq-mF-lower-bound-with-xxL}, only region of $m_F \gtrsim 1470 \, \textrm{GeV}$ is allowed for $x_L=-0.06$ to satisfy $s^\nu_L \leq 0.017$, i.e. the constraints on SM muon neutrino mixing with heavy neutral lepton as discussed in \cite{delAguila:2008pw}.
We can also see that the decoupling behavior of $a^\textrm{NP}_\mu$ with increasing $m_F$ is clearly manifested. This can be expected from Eq. \eqref{eq-xxL-as-input-Yukawa-expression} since fixed Yukawa couplings with larger $m_F$ would yield smaller mixings $s^\nu_L, s^\ell_L, s^\ell_R$ and smaller loop function values. More specifically, in the heavy region of $m_F$ we can have the following trending behavior of Eq. \eqref{eq-aNP-most-simplified-step-3} after utilizing the approximated form of loop functions in Eq. \eqref{eq-loop-function-class-2}
\begin{align}
a^\textrm{NP}_{\mu} &\approx 
\frac{\sqrt{2} m_{\mu}m_F}{4 \pi^2 m_\delta^2} (\lambda_L x_L) \,  \sqrt{- \sqrt{2} \, x_L  \frac{v_\delta}{m_F} }  \cdot [-2 \,(\frac{1}{4})\,(\frac{m_F^2}{m_\delta^2})^{-1}+ 5 \,(-\frac{1}{4})\,(\frac{m_F^2}{m_\delta^2})^{-1} ]   \nonumber \\
&\propto \frac{m_{\mu}}{m_F} \lambda_L (- x_L) \sqrt{(- x_L)  \frac{v_\delta}{m_F} } \, ,
\end{align}
in which we have taken  $x_L, \lambda_L$ to be real numbers to simplify the expression. One can easily see that $a^\textrm{NP}_{\mu}$ decreases with fixed $x_L,\lambda_L$ and increasing $m_F$.
In the right panel of Fig. \ref{fig-amuNP} we impose the $\{\lambda_L, s^\nu_L \}$ scheme described in Eq. \eqref{eq-snL-as-input-Yukawa-expression} as input parameters and take $\lambda_L=1, \, s^\nu_L=0.017$ as benchmark point. Aside from the observation that $a^\textrm{NP}_\mu$ can cover the central region of $\Delta a_\mu$ with $m_F\approx 1500 \, \textrm{GeV}$ in this parameter setup, noticeable chiral enhancement is manifested by enhanced $a^\textrm{NP}_\mu$ when $m_F$ increases.

\begin{figure}
    \centering
    \includegraphics[width= 0.482 \linewidth]{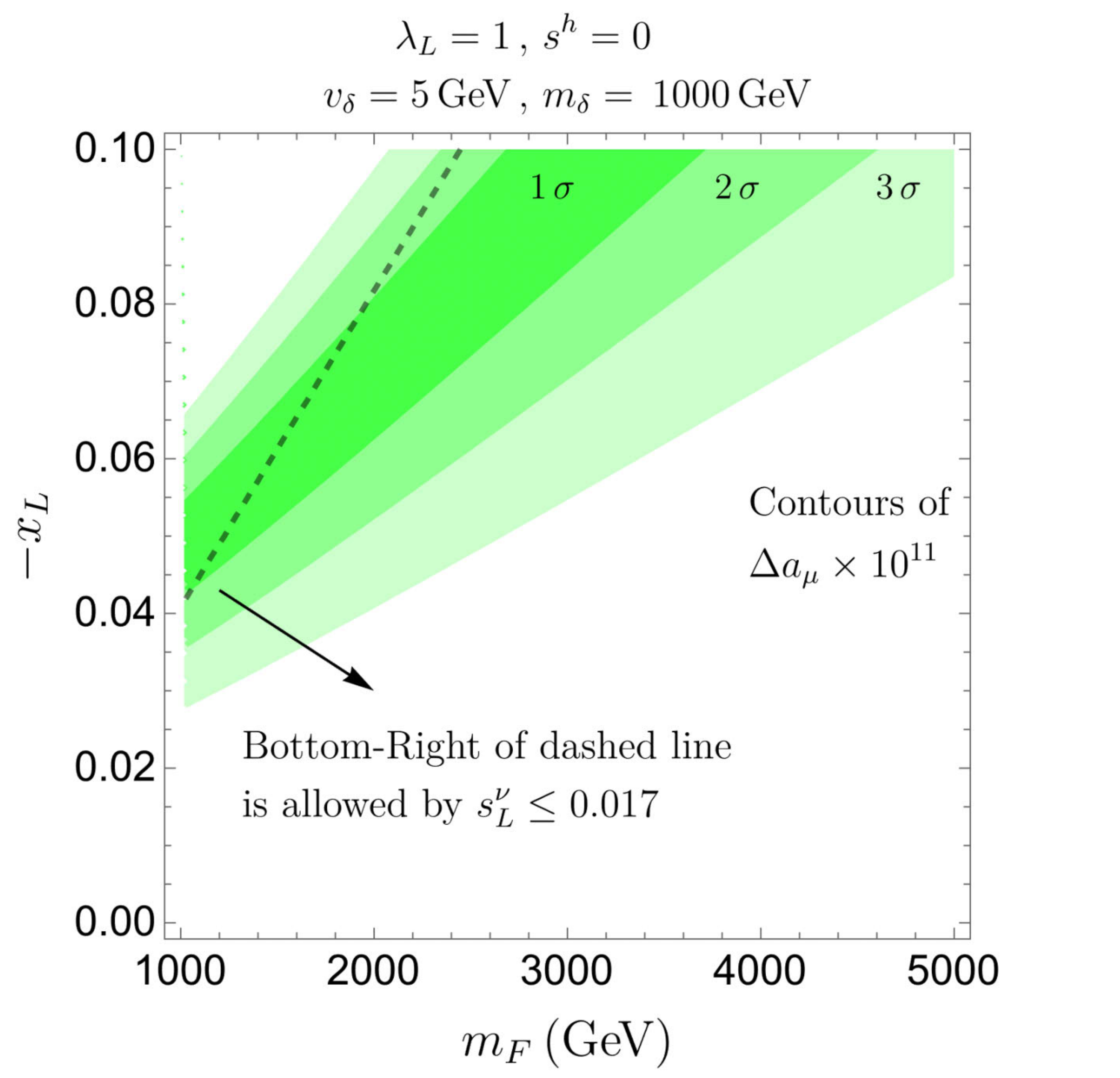}
    \includegraphics[width= 0.478 \linewidth]{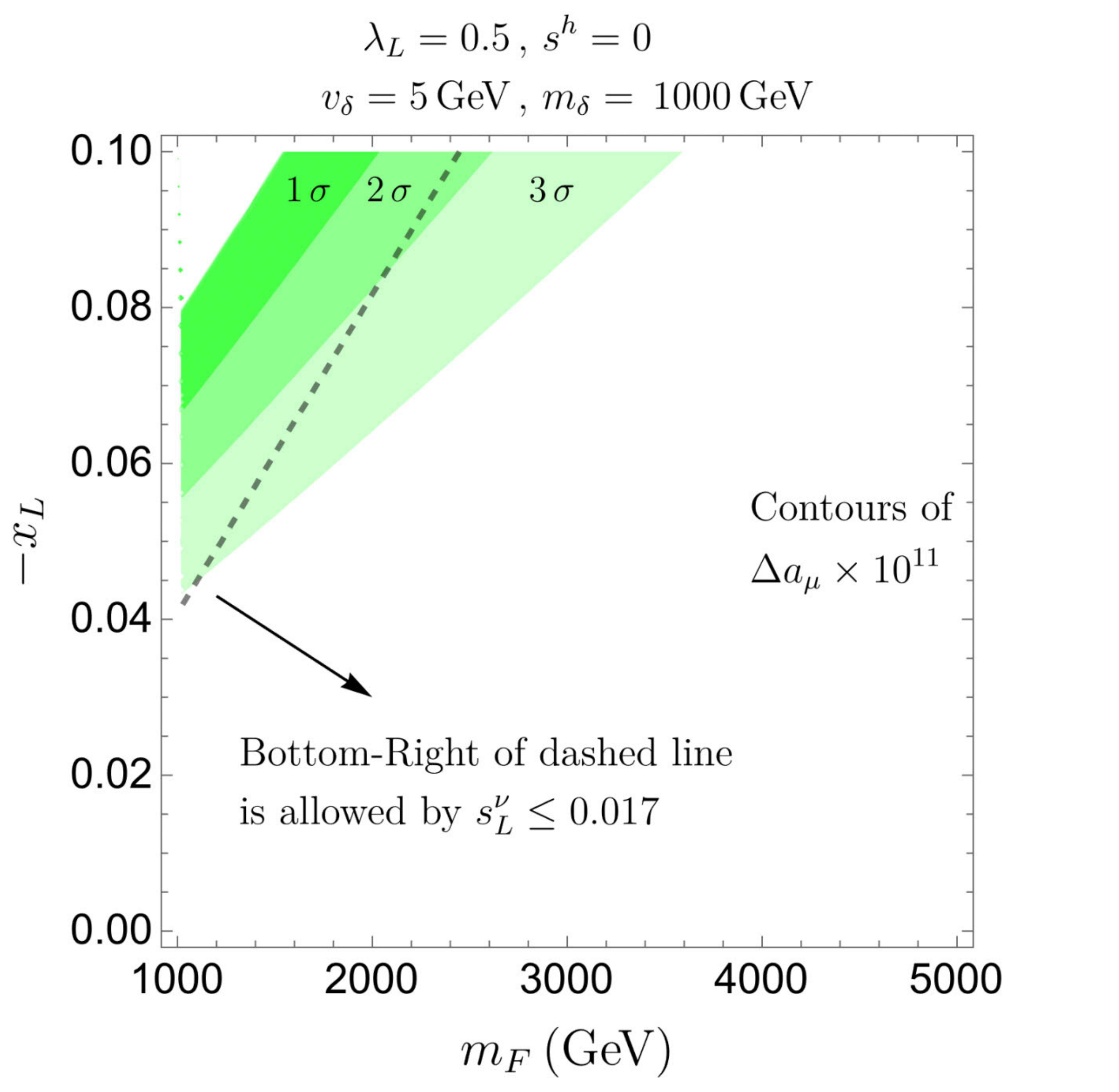}
    \caption{New physics contribution $a^\textrm{NP}_\mu$ predicted by our model on the plane of $m_F$ and $-x_L$ aiming at interpreting $\Delta a_\mu$ with dark, medium and light green color denoting the $1\sigma, \, 2\sigma, \, 3\sigma$ ranges. Dashed black line corresponds to the boundary of $s^\nu_L = 0.017$ discussed in \cite{delAguila:2008pw}, with respect to which the bottom-right region is allowed. {\bf Left}:$\lambda_L=1$. {\bf Right}: $\lambda_L=0.5$.}
    \label{fig-mF-vs-xL-amueF-contour}
\end{figure}

In Fig. \ref{fig-mF-vs-xL-amueF-contour} we show $a^\textrm{NP}_\mu$ on the plane of $m_F$ versus $-x_L$ in which the left (right) panel has $\lambda_L = 1$ ($\lambda_L = 0.5$). The green color of dark, medium and light opacity indicate the $1\sigma, \, 2\sigma, \, 3\sigma$ ranges of $\Delta a_\mu$. 
This figure can be view as the extended illustration of the left panel of Fig. \ref{fig-amuNP} by further allowing $x_L$ to vary. The dashed line denotes $s^\nu_L = 0.017$ with respect to which the bottom-right region is allowed. We can see that in the left panel, the region with $m_F \approx 1000 \, \textrm{GeV}$ and $x_L=-0.04$ can generate $a^\textrm{NP}_\mu$ on the edge of $1\, \sigma$ region of $\Delta a_\mu$. In the right panel with $\lambda_L = 0.5$, however, the same parameter space with $x_L=-0.04, \, m_F\approx 1000 \, \textrm{GeV}$ can only generate $a^\textrm{NP}_\mu$ on the edge of $3\, \sigma$ region and thus has little capability of explaining $\Delta a_\mu$.

\begin{figure}
    \centering
    \includegraphics[width= 0.482 \linewidth]{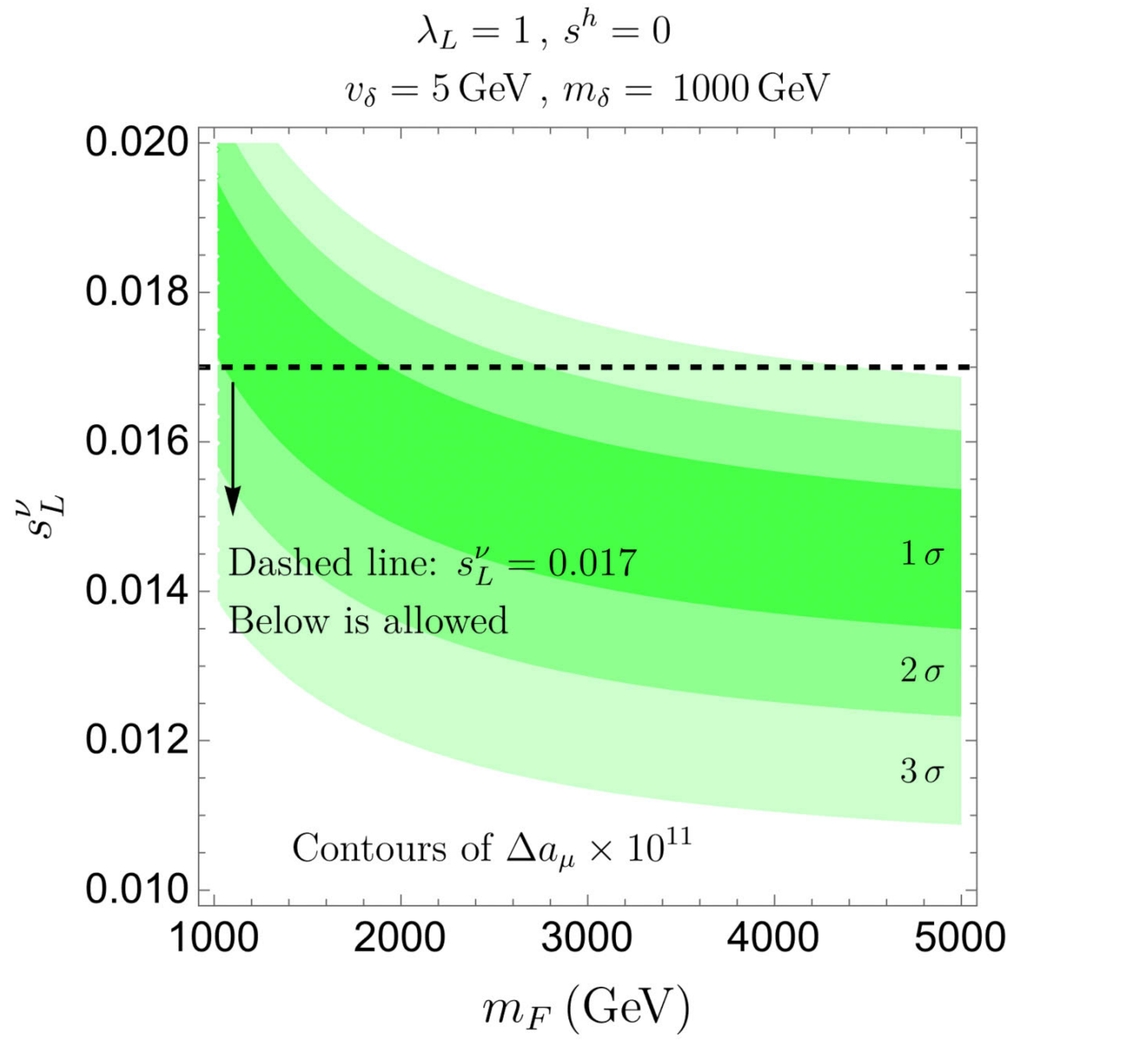}
    \includegraphics[width= 0.478 \linewidth]{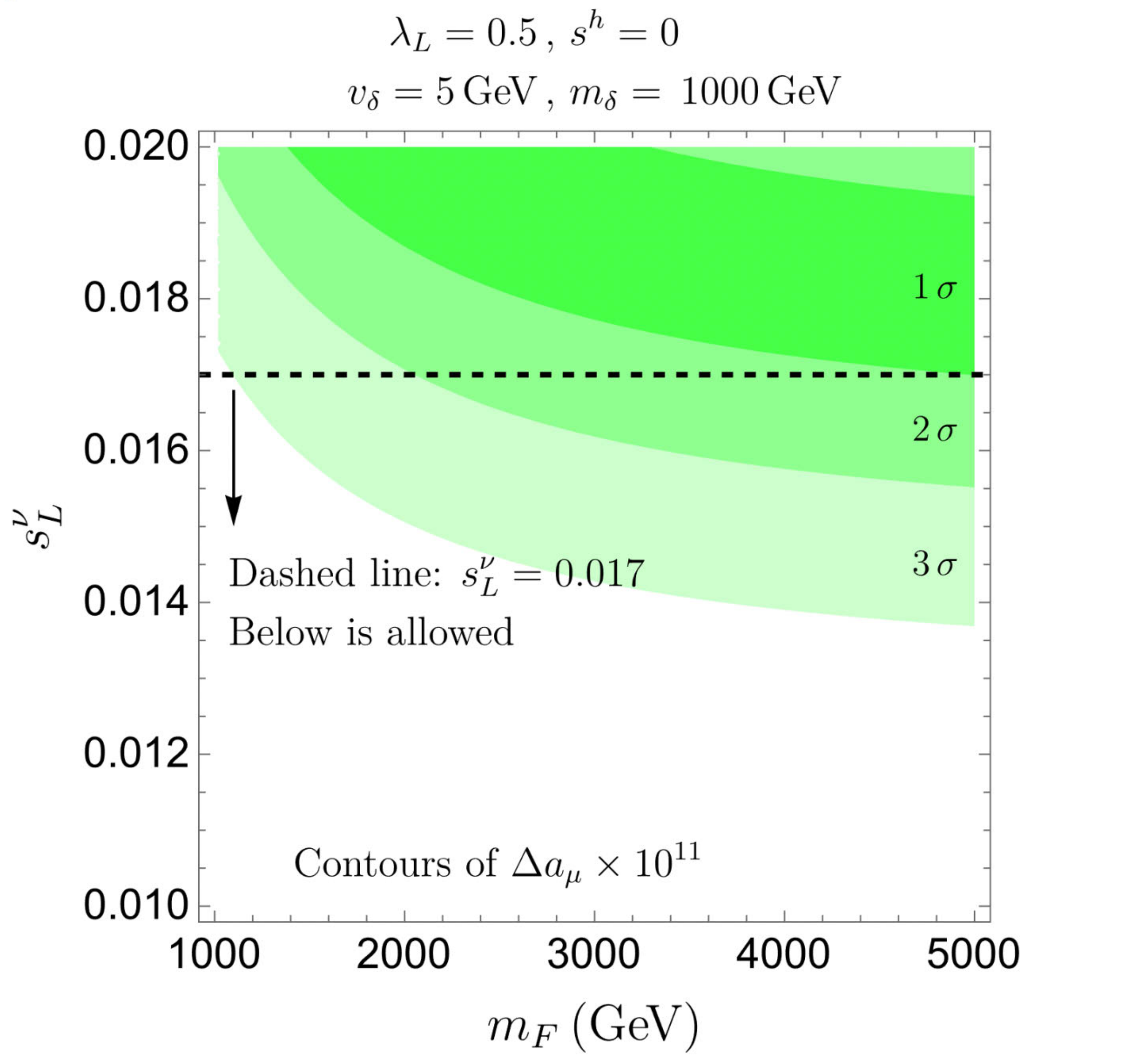}
    \caption{New physics contribution $a^\textrm{NP}_\mu$ predicted by our model in Eq. \eqref{eq-aNP} on the plane of $m_F$ and $s^\nu_L$ aiming at interpreting $\Delta a_\mu$ with dark, medium and light green color denoting the $1\sigma, \, 2\sigma, \, 3\sigma$ ranges. Dashed black line corresponds to the boundary of $s^\nu_L = 0.017$ discussed in \cite{delAguila:2008pw}, with respect to which the lower region is allowed. {\bf Left}: $\lambda_L=1$. {\bf Right}: $\lambda_L=0.5$.}
    \label{fig-mF-vs-snL-amueF-contour}
\end{figure}

In Fig. \ref{fig-mF-vs-snL-amueF-contour} we show $a^\textrm{NP}_\mu$ on the plane of $m_F$ versus $s^\nu_L$ in which the left (right) panel has $\lambda_L = 1$ ($\lambda_L = 0.5$).
This figure can be view as the extended illustration of the right panel of Fig. \ref{fig-amuNP} by further allowing $s^\nu_L$ to vary. The dashed line is $s^\nu_L = 0.017$ with respect to which the region below is allowed. We can see that in the left panel, the region with $s^\nu_L = 0.017$ and $1000 \, \textrm{GeV} \lesssim m_F \lesssim 2000\, \textrm{GeV}$ can generate $a^\textrm{NP}_\mu$ within the $1\, \sigma$ range of $\Delta a_\mu$. Region with smaller value of $s^\nu_L$ would require heavier $m_F$ and thus more significant chiral enhancement to achieve the same level of $a^\textrm{NP}_\mu$. The right panel, however, can only generate $a^\textrm{NP}_\mu$ on the edge of $1\, \sigma$ range of $\Delta a_\mu$ with quite heavy fermion mass $m_F \approx 5\, \textrm{TeV}$ with $s^\nu_L = 0.017$. We also checked that all Yukawa couplings in our model satisfy the requirement of perturbativity on the shown range of Fig. \ref{fig-mF-vs-snL-amueF-contour} which can be easily seen through Eq. \eqref{eq-snL-as-input-Yukawa-expression}.

\subsection{Phenomenological discussions}

In this section we briefly discuss the phenomenological aspects of our model.

Note that the scalar triplet and lepton triplet have been scrutinized in neutrino mass generation mechanisms as Type-II \cite{Magg:1980ut,Schechter:1980gr,Wetterich:1981bx,Lazarides:1980nt,Mohapatra:1980yp,Cheng:1980qt} and Type-III \cite{Foot:1988aq,Ma:2002pf} seesaw models, respectively. The relevant collier signals at LHC of these new physics triplets have been discussed in recent comprehensive reviews based on latest data collected at ATLAS \cite{ATLAS:2024fdw} and CMS \cite{CMS:2024bni} experiments. Generally speaking, lower bounds on masses around 1 TeV have been established for the new triplets, with specific assumptions on their production and decay modes.
Further discussions on the prospects of searches at future high luminosity LHC and proposed 100 TeV $pp$ collider can be found in \cite{Du:2018eaw, Banerjee:2024jwn, Bolton:2024thn} for scalar triplet and \cite{Das:2020uer, Das:2020gnt, Das:2023tna} for lepton triplet. Note that in this work we only concentrate on the physics of $(g-2)_\mu$ and do not require our model to produce the experimentally suggested texture of neutrino mass matrix. Instead, we require our model parameters to ensure the theoretically predicted neutrino mass to be negligibly small and close to zero. Aside from the neutrino physics, recently it has been found that the Type-II seesaw model can also address the problem of baryon asymmetry of the universe (see e.g. \cite{Barrie:2021mwi, Barrie:2022cub, Han:2022ssz, Han:2023kjg}).

We need to point out that the current bounds listed in \cite{ATLAS:2024fdw} and \cite{CMS:2024bni} cannot be applied directly to our model, given the fact that both scalar and lepton triplet exist in our model. Taking $m_F > m_\delta$ as a case example, which is the set up in our numerical analyses and is capable of explaining $\Delta a_\mu$, there is one main aspect differentiating our model from the exclusive Type-III seesaw signals. That is, after $F^0,F^{\pm}$ being singly or pairly produced, there are additional decay channels $F^0\rightarrow\ell^{\pm}\delta^{\mp}$, $F^-\rightarrow \ell^-\delta^0,\ell^-a^0,\ell^+\delta^{--}$, which do not exist in the existing search channels for Type-III seesaw mainly consisting of $F^0\rightarrow \nu h,\nu Z, \ell^\pm W^\mp$ and $F^\pm \rightarrow \ell^\pm h,\ell^\pm Z, \nu W^\pm$. 

To explain the above point, we first consider the partial decay widths of neutral heavy lepton $F^0$  given as follows.
\begin{align}
\Gamma(F^0\rightarrow \nu h)&\approx\frac{1}{16\pi m_F}\frac{m_F^2z_L^2}{4}\approx\frac{1}{16\pi m_F}\frac{m_F^4(s_L^{\nu})^2}{v^2},\nonumber\\
\Gamma(F^0\rightarrow \nu \delta^0)&\approx\frac{1}{8\pi m_F}(1-\frac{m_{\delta}^2}{m_F^2})m_F^2x_L^2(s_L^{\nu})^2\approx\frac{1}{16\pi m_F}(1-\frac{m_{\delta}^2}{m_F^2})\frac{m_F^4(s_L^{\nu})^6}{v_{\delta}^2},\nonumber\\
\Gamma(F^0\rightarrow \mu^{\pm}\delta^{\mp})&\approx\frac{1}{16\pi m_F}(1-\frac{m_{\delta}^2}{m_F^2})\frac{m_F^2\lambda_L^2}{2},\nonumber\\
\Gamma(F^0\rightarrow \nu Z)&\approx\frac{1}{16\pi m_F}\frac{m_F^4(s_L^{\nu})^2}{v^2},\nonumber\\
\Gamma(F^0\rightarrow \mu^{\pm}W^{\mp})&\approx\frac{1}{16\pi m_F}\frac{m_F^4(s_L^{\nu})^2}{v^2}.
\end{align}
Then, we find that
\begin{align}
&\Gamma(F^0\rightarrow \nu h):\Gamma(F^0\rightarrow \nu Z):\Gamma(F^0\rightarrow \mu^{\pm}W^{\mp})\approx1:1:1,\nonumber\\
&\Gamma(F^0\rightarrow \nu h):\Gamma(F^0\rightarrow \nu \delta^0):\Gamma(F^0\rightarrow \mu^{\pm}\delta^{\mp})\approx1:(1-\frac{m_{\delta}^2}{m_F^2})\frac{(s_L^{\nu})^4v^2}{v_{\delta}^2}:(1-\frac{m_{\delta}^2}{m_F^2})\frac{\lambda_L^2v^2}{2m_F^2(s_L^{\nu})^2}.
\end{align}
Taking $\lambda_L=1,v_{\delta}=5\ \mr{GeV},m_{\delta}=1\ \mr{TeV},m_F=2\ \mr{TeV},s_L^{\nu}=0.017$, the $\Gamma(F^0\rightarrow \nu \delta^0)$ is negligible. The other decay widths are estimated as 
\begin{align}
&\Gamma(F^0\rightarrow \nu h):\Gamma(F^0\rightarrow \nu Z):\Gamma(F^0\rightarrow \mu^{\pm}W^{\mp}):\Gamma(F^0\rightarrow \mu^{\pm}\delta^{\mp})\approx1:1:1:19.6.
\end{align}
We can see that $F^0\rightarrow \mu^{\pm}\delta^{\mp}$ is the dominating decay channel in the our $\Delta a_\mu$-favored parameter space.

Similarly, the partial decay widths of charged heavy lepton $F^-$ are computed as follows.
\begin{align}
\Gamma(F^-\rightarrow \mu^-h)&\approx\frac{1}{16\pi m_F}\frac{m_F^4(s_L^{\nu})^2}{v^2},\nonumber\\
\Gamma(F^-\rightarrow \mu^-\delta^0)&\approx\frac{1}{16\pi m_F}(1-\frac{m_{\delta}^2}{m_F^2})\frac{m_F^2\lambda_L^2}{4},\nonumber\\
\Gamma(F^-\rightarrow \nu\delta^-)&\approx\frac{1}{16\pi m_F}(1-\frac{m_{\delta}^2}{m_F^2})m_F^2(s_L^{\nu})^2[2x_L^2+\frac{1}{2}\lambda_L^2(s_R^{\ell})^2] \nonumber\\
&\approx\frac{1}{16\pi m_F}(1-\frac{m_{\delta}^2}{m_F^2})m_F^2(s_L^{\nu})^2[\frac{m_F^2(s_L^{\nu})^4}{v_{\delta}^2}+\frac{\lambda_L^4 v_{\delta}^2}{4m_F^2}],\nonumber\\
\Gamma(F^-\rightarrow \mu^+\delta^{--})&\approx\frac{1}{16\pi m_F}(1-\frac{m_{\delta}^2}{m_F^2})\frac{m_F^2\lambda_L^2}{2},\nonumber\\
\Gamma(F^-\rightarrow \mu^-Z)&\approx\frac{1}{16\pi m_F}\frac{m_F^4(s_L^{\nu})^2}{v^2},\nonumber\\
\Gamma(F^-\rightarrow \nu W^-)&\approx\frac{1}{16\pi m_F}\frac{2m_F^4(s_L^{\nu})^2}{v^2}.
\end{align}
Then, we find that
\begin{align}
&\Gamma(F^-\rightarrow \mu^-h):\Gamma(F^-\rightarrow \mu^-Z):\Gamma(F^-\rightarrow \nu W^-)\approx1:1:2,\nonumber\\
&\Gamma(F^-\rightarrow \mu^-h):\Gamma(F^-\rightarrow \mu^-\delta^0):\Gamma(F^-\rightarrow \nu\delta^-):\Gamma(F^-\rightarrow \mu^+\delta^{--})\nonumber\\
    &\approx1:(1-\frac{m_{\delta}^2}{m_F^2})\frac{\lambda_L^2v^2}{4m_F^2(s_L^{\nu})^2}:(1-\frac{m_{\delta}^2}{m_F^2})[\frac{(s_L^{\nu})^4v^2}{v_{\delta}^2}+\frac{\lambda_L^4v^2v_{\delta}^2}{4m_F^4}]:(1-\frac{m_{\delta}^2}{m_F^2})\frac{\lambda_L^2v^2}{2m_F^2(s_L^{\nu})^2}.
\end{align}
Taking $\lambda_L=1,v_{\delta}=5\ \mr{GeV},m_{\delta}=1\ \mr{TeV},m_F=2\ \mr{TeV},s_L^{\nu}=0.017$, the $\Gamma(F^-\rightarrow \nu\delta^-)$ is negligible. The other decay widths are estimated as 
\begin{align}
&\Gamma(F^-\rightarrow \mu^-h):\Gamma(F^-\rightarrow \mu^-Z):\Gamma(F^-\rightarrow \nu W^-):\Gamma(F^-\rightarrow \mu^-\delta^0):\Gamma(F^-\rightarrow \mu^+\delta^{--})\nonumber\\
&\approx1:1:2:9.8:19.6.
\end{align}
We can see that $F^-\rightarrow \mu^+\delta^{--}$ and $F^-\rightarrow \mu^-\delta^0$ are the leading and sub-leading contributions, respectively.

Now we continue to study the decay of $\delta^0,\delta^-,\delta^{--}$ as the decay products from $F^-\rightarrow \mu^+\delta^{--}$, $F^0\rightarrow \mu^{\pm}\delta^{\mp}$ and $F^-\rightarrow \mu^-\delta^0$. We have the following results
\begin{align}
&\Gamma(\delta^0\rightarrow \mu^+\mu^-):\Gamma(\delta^0\rightarrow W^+W^-)\sim1:\frac{2m_{\delta}^2v_{\delta}^2}{\lambda_L^2v^4(s_L^{\nu})^2}\nonumber\\
&\Gamma(\delta^-\rightarrow \mu^-\nu):\Gamma(\delta^-\rightarrow W^-Z)\sim1:\frac{2m_{\delta}^2v_{\delta}^2}{v^4[\lambda_L^2(s_L^{\nu})^2+(s_L^{\nu})^4m_F^2/v_{\delta}^2]}\nonumber\\
&\Gamma(\delta^{--}\rightarrow \mu^-\mu^-):\Gamma(\delta^{--}\rightarrow W^-W^-)\sim1:\frac{4m_{\delta}^2v_{\delta}^2}{v^4[\lambda_L^4v_{\delta}^2/m_F^2+(s_L^{\nu})^4m_F^2/v_{\delta}^2]}\nonumber\\.
\end{align}
Taking $\lambda_L=1,v_{\delta}=5\ \mr{GeV},m_{\delta}=1\ \mr{TeV},m_F=2\ \mr{TeV},s_L^{\nu}=0.017$, the decay ratios are estimated as
\begin{align}
&\Gamma(\delta^0\rightarrow \mu^+\mu^-):\Gamma(\delta^0\rightarrow W^+W^-)\sim1:47,\nonumber\\
&\Gamma(\delta^-\rightarrow \mu^-\nu):\Gamma(\delta^-\rightarrow W^-Z)\sim1:1,\nonumber\\
&\Gamma(\delta^{--}\rightarrow \mu^-\mu^-):\Gamma(\delta^{--}\rightarrow W^-W^-)\sim1:2.
\end{align}
We can see that $\delta^0, \delta^\pm, \delta^{\pm\pm}$ will further  decay into two gauge bosons $WW, WZ$ with dominating or sizable branching ratios \cite{Aoki:2011pz, Du:2018eaw}. Finally, $WW, WZ$ will decay into leptonic or hadronic states.

Therefore, in comparison to the existing searches for Type-III fermion triplets utilizing $2,3,4$ leptons accompanied by jets, our signal profile include even more leptons and jets. This increased particle multiplicity can enable more dedicated signal selection and background reduction and we leave the analysis for future works.

\section{Conclusions}\label{sec-conclusion}
In this work, we investigated the muon anomalous dipole moment $a_\mu$ in a model that extends the SM with a scalar triplet and a lepton triplet. We identify an important ingredient overlooked in previous studies, i.e. the Yukawa interaction involving the SM Higgs doublet and the newly introduced lepton triplet. This interaction is a four-dimensional operator and thus can naturally exist. We show that this Yukawa interaction can not only induce mass mixing between leptons in the SM and the new physics sector, but also provide an additional source of chiral flip to $a_\mu$. 
We find that there is still viable parameter space in this model to explain $\Delta a_\mu$, which is different from the observation reported in the existing literature. More specifically, while being consistent with the current data of neutrino mass, electroweak precision measurements and the perturbativity of couplings, our model can provide new physics contribution $a_\mu^\textrm{NP}$ to cover the central region of $\Delta a_\mu$ with new scalar and fermion mass as low as around TeV. This mass scale is allowed by the current collider searches for doubly charged scalars and the lepton triplet, and they can be tested at future high energy and/or high luminosity colliders.

\begin{acknowledgments}
 CH acknowledges supports from the National Key R{\&}D Program of China under grant 2023YFA1606100, the National Natural Science Foundation of China under Grant No. 12435005, the Sun Yat-Sen University Science Foundation, the Fundamental Research Funds for the Central Universities at Sun Yat-sen University under Grant No.\,24qnpy117, and the Key Laboratory of Particle Astrophysics and Cosmology (MOE) of Shanghai Jiao Tong University. SH is supported by the Basic Research Program of Shanxi Province (Grant No. 202403021222062) and by the startup research fund of Taiyuan University of Technology (Grant No. RY2400001554). SH would also like to acknowledge the hospitality of Center for High Energy Physics, Peking University, where he spent three months as a visitor. PW acknowledges support from Natural Science Foundation of Jiangsu Province (Grant No. BK20210201), Fundamental Research Funds for the Central Universities, Excellent Scholar Project of Southeast University (Class A), and the Big Data Computing Center of Southeast University.
\end{acknowledgments}
\section*{Appendix}
\begin{appendices}

\section{Details of Lagrangian terms and interactions}\label{appendix-sec-Details-of-Lagrangian-terms-and-interactions}

The Yukawa interactions can be written as 
\begin{equation}
\begin{aligned}\label{eq-lepton-mass-mixing-charged}
\mc{L}_\textrm{Yuk.}= & -\frac{1}{\sqrt{2}}y^{ij}h\overline{\ell_L^i}\ell_R^j-\frac{1}{\sqrt{2}}z_L^ih\overline{\ell_L^i}(F_L^+)^C-\frac{1}{2}z_L^ih\overline{\nu_L^i}(F_L^0)^C \\
& -\frac{1}{\sqrt{2}}\lambda_L^i\delta^0\overline{\ell_R^i}F_L^--\lambda_L^i\delta^-\overline{\ell_R^i}F_L^0-\lambda_L^i\delta^{--}\overline{\ell_R^i}F_L^+\\
&+\frac{1}{\sqrt{2}}x_L^{ij}\delta^0\overline{\nu_L^i}(\nu_L^j)^C-\frac{1}{\sqrt{2}}x_L^{ij}\delta^-\overline{\nu_L^i}(\ell_L^j)^C\\
&-\frac{1}{\sqrt{2}}x_L^{ij}\delta^-\overline{\ell_L^i}(\nu_L^j)^C-x_L^{ij}\delta^{--}\overline{\ell_L^i}(\ell_L^j)^C+\mathrm{h.c.} \, .
\end{aligned}
\end{equation}
In the above expression, the replacement $\delta^0 \to i \, a^0$ would generate the Yukawa interactions of $a^0$ and we do not write them explicitly for compactness. Moreover,
as will be discussed later, $G^0$ ($G^\pm$) will mix with $a^0$ ($\delta^\pm$) and yield mass eigenstate with zero mass, corresponding to the Goldstone bosons absorbed to be the longitudinal components of SM gauge bosons $Z$ ($W^\pm$) via the Higgs mechanism. Therefore, we also do not write terms of $G^0, G^\pm$ explicitly for compactness.

After the transformations in Eq. \eqref{eqn:model:1S1VLL:FL:rotation} we obtain the following Yukawa interactions in terms of mass eigenstates
\begin{equation}
\begin{aligned}
& \mc{L}_\textrm{Yuk.}= \enspace  \mc{L}_\textrm{Yuk.}^\textrm{I} + \mc{L}_\textrm{Yuk.}^\textrm{II} + \mc{L}_\textrm{Yuk.}^\textrm{III}  \,  ,
\end{aligned}
\end{equation}
\begin{equation}
\begin{aligned}
\mc{L}_\textrm{Yuk.}^\textrm{I}=&\quad\, \overline{\ell_L}\ell_R[(-\frac{y}{\sqrt{2}} c_L^\ell  c_R^\ell +\frac{z_L}{\sqrt{2}} c_L^\ell  s_R^\ell )h+\frac{1}{\sqrt{2}} \lambda_L^{\ast}  s_L^\ell  c_R^\ell \delta^0]  \\
&+\overline{\ell_L}(F_L^+)^C[(-\frac{y}{\sqrt{2}} c_L^\ell  s_R^\ell -\frac{z_L}{\sqrt{2}} c_L^\ell  c_R^\ell )h+\frac{1}{\sqrt{2}} \lambda_L^{\ast}  s_L^\ell  s_R^\ell \delta^0]\\
&+\overline{F_L^-}\ell_R[(-\frac{y}{\sqrt{2}} s_L^\ell  c_R^\ell +\frac{z_L}{\sqrt{2}} s_L^\ell  s_R^\ell )h-\frac{1}{\sqrt{2}} \lambda_L^{\ast}  c_L^\ell  c_R^\ell \delta^0]  \\
&+\overline{F_L^-}(F_L^+)^C[(-\frac{y}{\sqrt{2}} s_L^\ell  s_R^\ell -\frac{z_L}{\sqrt{2}} s_L^\ell  c_R^\ell )h-\frac{1}{\sqrt{2}} \lambda_L^{\ast}  c_L^\ell  s_R^\ell \delta^0] +\mathrm{h.c.} \, ,
\end{aligned}
\end{equation}
\begin{equation}
\begin{aligned}
\mc{L}_\textrm{Yuk.}^\textrm{II}=&\quad\, \delta^{--}[-x_L( c_L^\ell )^2\overline{\ell_L}(\ell_L)^C+\lambda_L s_R^\ell  c_R^\ell \overline{\ell_R}(\ell_R)^C]\\
&+\delta^{--}[-2x_L s_L^\ell  c_L^\ell \overline{\ell_L}(F_L^-)^C+\lambda_L\big(( s_R^\ell )^2-( c_R^\ell )^2\big)\overline{\ell_R}F_L^+]\\
&+\delta^{--}[-x_L( s_L^\ell )^2\overline{F_L^-}(F_L^-)^C-\lambda_L s_R^\ell  c_R^\ell \overline{(F_L^+)^C}F_L^+]\\
&+\delta^-[-\sqrt{2}x_Lc_L^{\nu} c_L^\ell \overline{\nu_L}(\ell_L)^C+\lambda_Ls_L^{\nu} c_R^\ell \overline{\ell_R}\nu_L]\\
&+\delta^-[-\sqrt{2}x_Lc_L^{\nu} s_L^\ell \overline{\nu_L}(F_L^-)^C-\lambda_Lc_L^{\nu} c_R^\ell \overline{\ell_R}F_L^0]\\
&+\delta^-[-\sqrt{2}x_Ls_L^{\nu} c_L^\ell \overline{F_L^0}(\ell_L)^C+\lambda_Ls_L^{\nu} s_R^\ell \overline{(F_L^+)^C}\nu_L]\\
&+\delta^-[-\sqrt{2}x_Ls_L^{\nu} s_L^\ell \overline{F_L^0}(F_L^-)^C-\lambda_Lc_L^{\nu} s_R^\ell \overline{(F_L^+)^C}F_L^0]+\mathrm{h.c.} \, ,
\end{aligned}
\end{equation}
\begin{equation}
\begin{aligned}
\mc{L}_\textrm{Yuk.}^\textrm{III}=&\quad\, \overline{\nu_L}(\nu_L)^C[\frac{1}{2}z_Ls_L^{\nu}c_L^{\nu}h+\frac{1}{\sqrt{2}}x_L(c_L^{\nu})^2\delta^0]\\
&+\overline{\nu_L}(F_L^0)^C[\frac{1}{2}z_L\big((s_L^{\nu})^2-(c_L^{\nu})^2\big)h+\sqrt{2}x_Ls_L^{\nu}c_L^{\nu}\delta^0]\\
&+\overline{F_L^0}(F_L^0)^C[-\frac{1}{2}z_Ls_L^{\nu}c_L^{\nu}h+\frac{1}{\sqrt{2}}x_L(s_L^{\nu})^2\delta^0]+\mathrm{h.c.} \, .
\end{aligned}
\end{equation}
Then, the $(g-2)_{\mu}$ related Yukawa interactions are collected and simplified as
\begin{equation}\label{eq-LYukawa-FL-simplified}
\begin{aligned}
\mc{L}_\textrm{Yuk.}=&\quad\, \overline{\ell_L}\ell_R[-\frac{ m_\ell }{{v_h}}( c_L^\ell )^2h+\frac{1}{\sqrt{2}} \lambda_L^{\ast}  s_L^\ell  c_R^\ell \delta^0]\\
&+\overline{\ell_L}(F_L^+)^C[-\frac{m_{F^\pm}}{{v_h}} s_L^\ell  c_L^\ell h+\frac{1}{\sqrt{2}} \lambda_L^{\ast}  s_L^\ell  s_R^\ell \delta^0]\\
&+\overline{F_L^-}\ell_R[-\frac{ m_\ell }{{v_h}} s_L^\ell  c_L^\ell h-\frac{1}{\sqrt{2}} \lambda_L^{\ast}  c_L^\ell  c_R^\ell \delta^0]\\
&+\delta^{--}[-x_L( c_L^\ell )^2\overline{\ell_L}(\ell_L)^C+\lambda_L s_R^\ell  c_R^\ell \overline{\ell_R}(\ell_R)^C]\\
&+\delta^{--}[-2x_L s_L^\ell  c_L^\ell \overline{\ell_L}(F_L^-)^C+\lambda_L\big(( s_R^\ell )^2-( c_R^\ell )^2\big)\overline{\ell_R}F_L^+]\\
&+\delta^-[-\sqrt{2}x_Lc_L^{\nu} c_L^\ell \overline{\nu_L}(\ell_L)^C+\lambda_Ls_L^{\nu} c_R^\ell \overline{\ell_R}\nu_L]\\
&+\delta^-[-\lambda_Lc_L^{\nu} c_R^\ell \overline{\ell_R}F_L^0-\sqrt{2}x_Ls_L^{\nu} c_L^\ell \overline{F_L^0}(\ell_L)^C]+\mathrm{h.c.} \, .
\end{aligned}
\end{equation}
Again, note that the replacement $\delta^0 \to i \, a^0$ in $\mc{L}_\textrm{Yuk.}$ would generate the Yukawa interactions of $a^0$ and we do not write them explicitly for compactness. After performing the rotations in Eq. \eqref{eq_scalar_rotation}, we can transform the above Yukawa interactions into the form in terms of mass eigenstates.

\section{Details of parameter setup}\label{appendix_sec_Details_of_parameter_setup}

\subsection{Relations of parameters}\label{appendix_subsec_Relation_of_parameters}
When diagonalizing the neutrino mass matrix in Eq. \eqref{eq-lepton-mass-mixing-neutral}, we have the following relations
\begin{align}\label{eqn:SFL:input:relations-neutral}
-\sqrt{2}x_Lv_{\delta}&=m_{\nu}(c_L^{\nu})^2+m_{F^0}(s_L^{\nu})^2 \, , \nonumber\\
z_L{v_h}&=2(m_{F^0}-m_{\nu})s_L^{\nu}c_L^{\nu} \, , \nonumber\\
M_F&=m_{\nu}(s_L^{\nu})^2+m_{F^0}(c_L^{\nu})^2 \, .
\end{align}
Similarly, when diagonalizing the charged lepton mass matrix in Eq. \eqref{eq-lepton-mass-mixing-neutral}, we have the following relations
\begin{align}\label{eqn:SFL:input:relations-charged}
\frac{1}{\sqrt{2}}y{v_h}&= m_\ell  c_L^\ell  c_R^\ell +m_{F^\pm} s_L^\ell  s_R^\ell  \, ,\quad\quad \frac{1}{\sqrt{2}}z_L{v_h}=- m_\ell  c_L^\ell  s_R^\ell +m_{F^\pm} s_L^\ell  c_R^\ell  \, ,\nonumber\\
\frac{1}{\sqrt{2}}  \lambda_L^{\ast} v_{\delta} &=- m_\ell  s_L^\ell  c_R^\ell +m_{F^\pm} c_L^\ell  s_R^\ell  \, ,\quad\quad\quad\,\, M_F= m_\ell  s_L^\ell  s_R^\ell +m_{F^\pm} c_L^\ell  c_R^\ell \, .
\end{align}
Given that there are different equalities on $z_L v_h$ and $M_F$ in Eqs. \eqref{eqn:SFL:input:relations-neutral} and \eqref{eqn:SFL:input:relations-charged}, we can have the following identities
\begin{equation}
\begin{aligned}\label{eqn:SFL:input:zLML}
m_{F^0}&=\frac{1}{(c_L^{\nu})^2} \left( m_\ell  s_L^\ell  s_R^\ell +m_{F^\pm} c_L^\ell  c_R^\ell -m_{\nu}(s_L^{\nu})^2 \right),\\
- m_\ell  c_L^\ell  s_R^\ell +m_{F^\pm} s_L^\ell  c_R^\ell&=\sqrt{2}\frac{s_L^{\nu}}{c_L^{\nu}} \left( m_\ell  s_L^\ell  s_R^\ell +m_{F^\pm} c_L^\ell  c_R^\ell -m_{\nu} \right) \, .
\end{aligned}
\end{equation}
Considering that the current status of SM neutrino mass measurements from cosmology suggest $\sum_{\nu} m_\nu < 1 \, \textrm{eV}$ \cite{SajjadAthar:2021prg,ParticleDataGroup:2022pth}, we would  apply $m_\nu\to0$ as a constraint on the model parameters in the above relations. As for the SM charged lepton masses $m_\ell$ we utilize the non-zero value provided by Particle Data Group \cite{ParticleDataGroup:2022pth} in numerical calculations. However, one can still apply $ m_\ell \to0$ to obtain more compact analytical relations. For Eq. \eqref{eqn:SFL:input:relations-neutral} we have
\begin{align}\label{eqn:SFL:input:relationsapp-neutral}
-\sqrt{2} x_Lv_{\delta}\approx m_{F^0}(s_L^{\nu})^2,\quad z_L{v_h}\approx2m_{F^0}s_L^{\nu}c_L^{\nu},\quad M_F\approx m_{F^0}(c_L^{\nu})^2 \, ,
\end{align}
and for Eq. \eqref{eqn:SFL:input:relations-charged} we have
\begin{align}\label{eqn:SFL:input:relationsapp-charged}
\frac{1}{\sqrt{2}}y{v_h}&\approx m_{F^\pm} s_L^\ell  s_R^\ell  \, ,\quad\quad \,\,\frac{1}{\sqrt{2}}z_L{v_h}\approx m_{F^\pm} s_L^\ell  c_R^\ell  \, ,\nonumber\\
\frac{1}{\sqrt{2}}  \lambda_L^{\ast} v_{\delta} &\approx m_{F^\pm} c_L^\ell  s_R^\ell  \, ,\quad\quad\quad\quad\,\, M_F\approx m_{F^\pm} c_L^\ell  c_R^\ell \, .
\end{align}
Similarly, applying $m_{\nu}\rightarrow0$ and $m_\ell \to 0$ would reduce Eq. \eqref{eqn:SFL:input:zLML} to the following simple form
\begin{align}\label{eq-remove-two-extra-dof}
&m_{F^0}\approx m_{F^\pm} \frac{c_L^\ell  c_R^\ell }{(c_L^{\nu})^2} \, ,\quad \tan \theta_L^\ell \approx\sqrt{2}\tan\theta_L^{\nu} \, .
\end{align}
Based on the above discussion we have the following consideration on parameter setup.
\begin{itemize}
    \item $s_L^{\nu}$ characterizing the mixing of SM neutrino $\nu$ with $F^0$ is constrained by the electroweak precision measurements, for the muon flavor to be \cite{delAguila:2008pw},
\begin{align}\label{eq-snL-exp-constraint}
s^\nu_L \leq 0.017 \, ,
\end{align}
which can yield the following approximations with small mixing angle in the LH lepton sector
\begin{align}\label{eq-small-mixing-angle-LH}
s_L^\ell  \approx \sqrt{2} \, s_L^{\nu}  \sim  \mathcal{O}(10^{-2}) \, , \quad c^\nu_L \approx c^\ell_L \approx 1 \, , \quad m_{F^0}\approx m_{F^\pm} \, c^\ell_R \, .
\end{align}
This implies that nearly linear correlation exists between the above physical quantities which should be took into consideration when choosing independent input parameters.
    \item $s_R^{\ell}$ characterizing the mixing of RH component of SM charged lepton $\ell$ with $F^-$ can be solved from Eqs. \eqref{eqn:SFL:input:relationsapp-neutral} and \eqref{eqn:SFL:input:relationsapp-charged}. In the parameter regions chosen for our numerical calculation it turns out to be the following \footnote{Note that in other models the suppression of $s^\ell_R$ compared to $s^\ell_L$ can be different from our model. Taking Type-III seesaw model as an example, the suppression is $\sim m_\ell/M_F$ \cite{delAguila:2008pw, Abada:2008ea}.}
\begin{align}\label{eq-seR-compared-to-seL}
s^\ell_R \sim \mathcal{O}(10^{-1}) \, s^\ell_L \, ,
\end{align}
 which means that the small angle approximation also holds for $\theta_R$ (see Eqs. \eqref{eq-xxL-as-input-Yukawa-expression} and \eqref{eq-snL-as-input-Yukawa-expression} for more details). Therefore, in our model we have
\begin{align}\label{eq-small-mixing-angle-RH}
s^\ell_R  \sim  \mathcal{O}(10^{-3}) \, , \quad c^\ell_R \approx 1 \, , \quad m_{F^0}\approx m_{F^\pm} \, .
\end{align}
    \item 
    As the mass gap $\Delta m_F\equiv m_{F^\pm}-m_{F^0}$ in Eq. \eqref{eq-remove-two-extra-dof} is negligibly small, $\Delta m_F$ does not alter the main decay signals of the heavy leptons $F^\pm, F^0$ as discussed in \cite{CMS:2019hsm, ATLAS:2022yhd}. Therefore, we can simply impose the similar lower bound of mass to be
    \begin{align}\label{eq-mFpm-mF0-lower-bound}
        m_{F^\pm} \, , \, m_{F^0} \, \gtrsim 1000 \, {\rm GeV} \, .
    \end{align}   
\end{itemize}

\subsection{Input parameters}\label{appendix_subsec_Input_parameters}
Now we determine the physically reasonable choice of input parameters of our model.

In the scalar sector shown in Eq. \eqref{eq-Lag-VHS}, despite the rich parameters and phenomenology about $V(H,S)$ (see e.g. \cite{Du:2018eaw} for collider signal searches), the physical quantities most relevant to $(g-2)_\mu$ are
\begin{align}
v_{\delta} \, , \, v_{h} \, ,  \qquad m_h \, , \, m_{\delta^0} \, , \, m_{a^0} \, , \, m_{\delta^-} \, , \, m_{\delta^{--}}   \, , \qquad s^h \, , \, s^a \, , \, s^G \, ,
\end{align}
which are internally related \cite{Arhrib:2011uy}. 
Our requirements and simplifications include the following ones.
\begin{itemize}
    \item As for the two vevs $v_h, v_\delta$, we require them to satisfy \cite{Gunion:1989we, Aoki:2012jj} 
    \begin{align}\label{eq-vSM-value}
        v=\sqrt{v_{h}^2+2v_{\delta}^2}\approx 246 \, {\rm GeV} \, ,
    \end{align}   
    to be consistent with the electroweak precision measurement \cite{ParticleDataGroup:2022pth} and
    \begin{align}\label{eq-mh-value}
        m_h=125 \, {\rm GeV} \, ,
    \end{align}    
    to match the mass of the SM-like Higgs boson discovered at the Large Hadron Collider (LHC) \cite{ATLAS:2012yve,CMS:2012qbp,ParticleDataGroup:2022pth}.
    To avoid the constraints $ v_{\delta}\lesssim 5 \, {\rm GeV}$ from the electroweak precision observables \cite{Dev:2017ouk}, in our numerical calculation we choose
        \begin{align}\label{eq-vdelta-value}
        v_{\delta} = 5 \, {\rm GeV} \, , \, v_h \approx 246 \, {\rm GeV} \, .
    \end{align}  
    \item As for the masses of heavy scalars in the new physics sector, we require that the following approximations hold well
    \begin{align}\label{eq-MS-relation}
    M_S  \approx m_{\delta^0} \approx m_{a^0} \approx m_{\delta^-} \approx m_{\delta^{--}} \, ,
    \end{align}
    which can be properly realized under the conditions of $v_\delta \ll v_h$ and $m_h \ll M_S$ \cite{Arhrib:2011uy}. Hereafter we would use $m_\delta$ as the notation to denote 
    \begin{align}\label{eq-notation-mass-delta}
    m_\delta \equiv m_{\delta^0} \approx m_{a^0} \approx m_{\delta^-} \approx m_{\delta^{--}} \, .
    \end{align}
    Considering that the current searches for doubly charged scalar sets a lower limit of mass to be around $1 \, {\rm TeV}$ assuming decaying to SM leptons \cite{CMS:2017pet, ATLAS:2022pbd}, we take the following benchmark throughout this work
    \begin{align}\label{eq-mdelta-value}
        m_\delta = 1000 \, {\rm GeV} \, .
    \end{align}   
    \item As for the mixing $s^h,s^a,s^G$ in the scalar sector, we first have the following requirement to safely pass the current constraints from the study of Higgs data \cite{ATLAS:2021vrm,CMS:2020gsy}
    \begin{align}\label{eq-sh-value}
        |s^h| \lesssim \mathcal{O}(0.1) \, ,
    \end{align}  
    which can be properly realized due to the rich parameter space of $V(H,S)$. Moreover, our choice of $v_\delta, v_h$ in Eq. \eqref{eq-vSM-value} and \eqref{eq-vdelta-value} also results in \cite{Arhrib:2011uy}
    \begin{align}\label{eq-sa-sG-value}
        s^a \approx \tan\theta^a = \frac{2 \, v_\delta}{v_h} \sim \mathcal{O}(10^{-2}) \, , \quad  s^G \approx \tan\theta^G = \frac{1}{\sqrt{2}} \tan\theta^a  \sim \mathcal{O}(10^{-2}) \, ,
    \end{align} 
    which make the mixing $s^h,s^a,s^G$ all in the small value region. While keeping $s^h, s^a, s^G$ in the analytical results, we will simply impose the following simplifications in numerical calculation which make negligible difference to the results 
        \begin{align}\label{eq-sh-value}
            s^h=s^a=s^G=0 \, .
        \end{align}
\end{itemize}

\subsection{Perturbativity requirement}\label{appendix_subsec_Perturbativity_requirement}
Based on the discussion in the above it is easy to study the perturbativity behavior of the Yukawa couplings in our model.
\begin{itemize}
\item In the $\{\lambda_L, x_L\}$ scheme, we have the following approximations from Eq. \eqref{eq-xxL-as-input-Yukawa-expression}
\begin{align}
z_L \sim \sqrt{\frac{-x_L}{10^4}\,\frac{m_F}{\rm GeV}} \, , \quad y \sim \sqrt{\frac{-x_L}{100}\,\frac{\rm GeV}{m_F}} \, .
\end{align}
In this work we focus on $1 \, {\rm TeV} \lesssim m_F \lesssim 5 \, {\rm TeV}$ and $|x_L|\lesssim \mathcal{O}(10^{-1})$, thus the requirement of perturbativity $z_L, |x_L|, y < \mathcal{O}(1)$ can be easily satisfied. Note that Eq. \eqref{eq-snL-exp-constraint} and Eq. \eqref{eq-xxL-as-input-Yukawa-expression} also imply a lower bound of $m_F$ satisfying
\begin{align}\label{eq-mF-lower-bound-with-xxL}
m_F \gtrsim (-x_L) \frac{\sqrt{2} \, v_\delta}{(0.017)^2}  \approx 25 \,|x_L|\, {\rm TeV} \, ,
\end{align}
which will be manifested in our numerical results discussed later (see e.g. Fig. \ref{fig-amuNP}).
\item In the $\{\lambda_L, s_L^{\nu}\}$ scheme, we have the following approximations from Eq. \eqref{eq-snL-as-input-Yukawa-expression}
\begin{align}\label{eq-snL-lamL-as-input-perturbativity-bound}
z_L \sim \frac{1}{100}\,\frac{m_F}{\rm GeV}\, s^\nu_L \, , \quad   x_L \sim - \frac{1}{10}\,\frac{m_F}{\rm GeV}  (s^\nu_L)^2 \, , \quad y \sim \frac{1}{10} s^\nu_L \, .
\end{align}
Given $s^\nu_L \leq 0.017$ indicated in Eq. \eqref{eq-snL-exp-constraint}, we can see that the requirement of perturbativity $z_L, |x_L|, y < \mathcal{O}(1)$ can also be easily satisfied for the mass region $1 \, {\rm TeV} \lesssim m_F \lesssim 5 \, {\rm TeV}$ in our discussion.
\end{itemize}

\section{Details of analytical results}\label{appendix-sec-analytica-results}

The explicit form of $a_{\mu}^{\ell,\,\textrm{total}}$ originating from the left panel of Fig. \ref{fig-muongm2VLL_Feynman} is
\begin{align}\label{eq-mF-vs-xL-aell}
a_{\mu}^{\ell,\,\textrm{total}}=a_{\mu}^{\ell,\,h}+a_{\mu}^{\ell,\,\delta^0}+a_{\mu}^{\ell,\,a^0}+a_{\mu}^{\ell,\,\delta^{--}} \, ,
\end{align}
in which the neutral CP-even scalars $h,\delta^0$ contribute as
\begin{align}\label{eq-mF-vs-xL-aell-neutral-CP-even}
a_{\mu}^{\ell,\,h}&=\quad\frac{m_{\mu}^2}{4\pi^2m_h^2}[|\frac{m_{\mu}}{v_h}( c_L^\ell )^2c^h+\frac{1}{\sqrt{2}} \lambda_L s_L^\ell  c_R^\ell s^h|^2 -\frac{m_{\mu}^2}{v^2}]\cdot F_{LL}^{f,1}(\frac{m_{\mu}^2}{m_h^2}) \nonumber\\
&\quad +\frac{m_{\mu}^2}{4\pi^2m_h^2} \textrm{Re}[ (\frac{m_{\mu}}{v_h}( c_L^\ell )^2c^h+\frac{1}{\sqrt{2}} \lambda_L  s_L^\ell  c_R^\ell s^h)^2 - \frac{m_{\mu}^2}{v^2}] \cdot F_{LR}^{f,1}(\frac{m_{\mu}^2}{m_h^2}) \, , \nonumber\\
a_{\mu}^{\ell,\,\delta^0}&=\quad\frac{m_{\mu}^2}{4\pi^2m_{\delta^0}^2} |\frac{m_{\mu}}{v_h}( c_L^\ell )^2s^h-\frac{1}{\sqrt{2}} \lambda_L  s_L^\ell  c_R^\ell c^h|^2 \cdot F_{LL}^{f,1}(\frac{m_{\mu}^2}{m_{\delta^0}^2})\nonumber\\
&\quad +\frac{m_{\mu}^2}{4\pi^2m_{\delta^0}^2} \textrm{Re}[ (\frac{m_{\mu}}{v_h}( c_L^\ell )^2s^h-\frac{1}{\sqrt{2}} \lambda_L  s_L^\ell  c_R^\ell c^h)^2] \cdot F_{LR}^{f,1}(\frac{m_{\mu}^2}{m_{\delta^0}^2}) \, ,
\end{align}
and the neutral CP-odd scalar $a^0$ contributes as
\begin{align}\label{eq-mF-vs-xL-aell-neutral-CP-odd}
a_{\mu}^{\ell,\,a^0}&=\quad\frac{m_{\mu}^2}{8\pi^2m_{a^0}^2} |\lambda_L|^2 (c^a s^\ell_L c^\ell_R )^2  \cdot F_{LL}^{f,1}(\frac{m_{\mu}^2}{m_{a^0}^2})\nonumber\\
&\quad +\frac{m_{\mu}^2}{8\pi^2m_{a^0}^2} \textrm{Re}[-(\lambda_L c^a s^\ell_L c^\ell_R)^2 ] \cdot F_{LR}^{f,1}(\frac{m_{\mu}^2}{m_{a^0}^2}) \, , 
\end{align}
and the charged scalar $\delta^{--}$ contributes as
\begin{align}\label{eq-mF-vs-xL-aell-charged}
a_{\mu}^{\ell,\,\delta^{--}} &=\quad \frac{m_{\mu}^2}{8\pi^2m_{\delta^{--}}^2}[|x_L |^2 (2 c_L^\ell )^2 +| \lambda_L |^2 (2 s_R^\ell  c_R^\ell )^2 ]\cdot[-F_{LL}^{f,1}(\frac{m_{\mu}^2}{m_{\delta^{--}}^2})+2F_{LL}^{S,1}(\frac{m_{\mu}^2}{m_{\delta^{--}}^2})]\nonumber\\
&\quad  +\frac{m_{\mu}^2}{4\pi^2m_{\delta^{--}}^2} \textrm{Re}[\big(- 2 x_L ( c_L^\ell )^2 \big) ( 2 \lambda_L^{\ast} s_R^\ell  c_R^\ell )]  \cdot[-F_{LR}^{f,1}(\frac{m_{\mu}^2}{m_{\delta^{--}}^2})+2F_{LR}^{S,1}(\frac{m_{\mu}^2}{m_{\delta^{--}}^2})] \, .
\end{align}
Note that we have subtracted the SM Higgs contribution from $a_{\mu}^{\ell,\,h}$ to meet the definition of $a^\textrm{NP}$, i.e. the contribution exclusively generated from new physics sector. Moreover, the symmetry factor from the coupling of $\bar{\ell}\ell^C \delta^{--}$ has been properly considered in $a_{\mu}^{\ell,\,\delta^{--}}$ as pointed out in Ref. \cite{Queiroz:2014zfa}.

The explicit form of $a_{\mu}^{F}$ originating from the middle panel of Fig. \ref{fig-muongm2VLL_Feynman} is
\begin{align}\label{eq-mF-vs-xL-amueF}
a_{\mu}^{F,\,\textrm{total}}=a_{\mu}^{F,\,h}+a_{\mu}^{F,\,\delta^0}+a_{\mu}^{F,\,a^0}+a_{\mu}^{F,\,\delta^{--}}+a_{\mu}^{F,\,\delta^{-}} \, ,
\end{align}
in which the neutral CP-even scalars $h,\delta^0$ contribute as
\begin{align}\label{eq-mF-vs-xL-amueF-neutral-CP-even}
a_{\mu}^{F,\,h}&=\quad\frac{m_{\mu}^2}{8\pi^2m_h^2}[|\frac{m_{F^{\pm}}}{v_h} s_L^\ell  c_L^\ell c^h+\frac{1}{\sqrt{2}}\lambda_L s_L^\ell  s_R^\ell s^h|^2+|\frac{m_{\mu}}{v_h} s_L^\ell  c_L^\ell c^h-\frac{1}{\sqrt{2}}\lambda_L c_L^\ell  c_R^\ell s^h|^2]\cdot F_{LL}^{f,2}(\frac{m_{F^{\pm}}^2}{m_h^2}) \nonumber\\
&\quad +\frac{m_{\mu}m_{F^{\pm}}}{4\pi^2m_h^2} \textrm{Re}[ (\frac{m_{F^{\pm}}}{v_h} s_L^\ell  c_L^\ell c^h+\frac{1}{\sqrt{2}}\lambda_L s_L^\ell  s_R^\ell s^h)(\frac{m_{\mu}}{v_h} s_L^\ell  c_L^\ell c^h-\frac{1}{\sqrt{2}}\lambda_L c_L^\ell  c_R^\ell s^h)] \cdot F_{LR}^{f,2}(\frac{m_{F^{\pm}}^2}{m_h^2}) \, , \nonumber\\
a_{\mu}^{F,\,\delta^0}&=\quad\frac{m_{\mu}^2}{8\pi^2m_{\delta^0}^2}[|\frac{m_{F^{\pm}}}{v_h} s_L^\ell  c_L^\ell s^h -\frac{1}{\sqrt{2}}\lambda_L s_L^\ell  s_R^\ell c^h|^2+|\frac{m_{\mu}}{v_h} s_L^\ell  c_L^\ell s^h+\frac{1}{\sqrt{2}}\lambda_L c_L^\ell  c_R^\ell c^h|^2]\cdot F_{LL}^{f,2}(\frac{m_{F^{\pm}}^2}{m_{\delta^0}^2})\nonumber\\
&\quad +\frac{m_{\mu}m_{F^{\pm}}}{4\pi^2m_{\delta^0}^2} \textrm{Re}[ (\frac{m_{F^{\pm}}}{v_h} s_L^\ell  c_L^\ell s^h-\frac{1}{\sqrt{2}}\lambda_L s_L^\ell  s_R^\ell c^h)(\frac{m_{\mu}}{v_h} s_L^\ell  c_L^\ell s^h+\frac{1}{\sqrt{2}}\lambda_L c_L^\ell  c_R^\ell c^h) ] \cdot F_{LR}^{f,2}(\frac{m_{F^{\pm}}^2}{m_{\delta^0}^2}) \, ,
\end{align}
and the neutral CP-odd scalar $a^0$ contributes as
\begin{align}\label{eq-mF-vs-xL-amueF-neutral-CP-odd}
a_{\mu}^{F,\,a^0}&=\quad\frac{m_{\mu}^2}{8\pi^2m_{a^0}^2}[  \frac{1}{2}  |\lambda_L|^2 (c^a s^\ell_L s^\ell_R )^2  + \frac{1}{2}  | \lambda_L |^2 (c^a c^\ell_L c^\ell_R )^2]\cdot F_{LL}^{f,2}(\frac{m_{F^{\pm}}^2}{m_{a^0}^2})\nonumber\\
&\quad +\frac{m_{\mu}m_{F^{\pm}}}{4\pi^2m_{a^0}^2} \textrm{Re}[(\frac{1}{\sqrt{2}} \lambda_L c^a s^\ell_L s^\ell_R)  (\frac{1}{\sqrt{2}}\lambda_L c^a c^\ell_L c^\ell_R)  ] \cdot F_{LR}^{f,2}(\frac{m_{F^{\pm}}^2}{m_{a^0}^2}) \, , 
\end{align}
and the charged scalars $\delta^-,\delta^{--}$ contribute as
\begin{align}\label{eq-mF-vs-xL-amueF-charged}
a_{\mu}^{F,\,\delta^{--}}&=\quad\frac{m_{\mu}^2}{8\pi^2m_{\delta^{--}}^2}[4|x_L |^2 (s_L^\ell  c_L^\ell)^2 +|\lambda_L|^2 \big( ( c_R^\ell )^2-( s_R^\ell )^2 \big)^2]\cdot[-F_{LL}^{f,2}(\frac{m_{F^{\pm}}^2}{m_{\delta^{--}}^2})+2F_{LL}^{S,2}(\frac{m_{F^{\pm}}^2}{m_{\delta^{--}}^2})]\nonumber\\
&\quad +\frac{m_{\mu}m_{F^{\pm}}}{4\pi^2m_{\delta^{--}}^2} \textrm{Re}[ (2x_L s_L^\ell  c_L^\ell  ) \big( \lambda_L^{\ast} (( c_R^\ell )^2-( s_R^\ell )^2 ) \big)  ] \cdot[-F_{LR}^{f,2}(\frac{m_{F^{\pm}}^2}{m_{\delta^{--}}^2})+2F_{LR}^{S,2}(\frac{m_{F^{\pm}}^2}{m_{\delta^{--}}^2})] \, , \nonumber\\
a_{\mu}^{F,\,\delta^{-}}&= \quad \frac{m_{\mu}^2}{8\pi^2m_{\delta^-}^2}[2|x_L|^2( s_L^{\nu}  c_L^\ell c^G)^2+|\lambda_L|^2( c_L^{\nu}  c_R^\ell c^G )^2]\cdot F_{LL}^{S,2}(\frac{m_{F^0}^2}{m_{\delta^{-}}^2})\nonumber\\
&\quad +\frac{m_{\mu}m_{F^0}}{4\pi^2m_{\delta^-}^2} \textrm{Re} [(\sqrt{2} x_L  s_L^{\nu} c_L^\ell c^G )(  \lambda_L^{\ast}  c_L^{\nu} c_R^\ell c^G )] \cdot F_{LR}^{S,2}(\frac{m_{F^0}^2}{m_{\delta^-}^2}) \, .
\end{align}

The explicit form of $a_{\mu}^{\nu}$ originating from the right panel of Fig. \ref{fig-muongm2VLL_Feynman} is
\begin{align}
a_{\mu}^{\nu,\,\textrm{total}}&=\quad \frac{m_{\mu}^2}{8\pi^2m_{\delta^-}^2} [2|x_L|^2(c_L^{\nu} c_L^\ell  c^G )^2+|\lambda_L|^2(s_L^{\nu}  c_R^\ell c^G )^2]\cdot F_{LL}^{S,3}(\frac{m_{\mu}^2}{m_{\delta^-}^2})\nonumber\\
&\quad + \frac{m_{\mu} m_\nu}{4\pi^2m_{\delta^-}^2} \textrm{Re}[(-\sqrt{2} x_L  c_L^{\nu} c_L^\ell c^G ) ( \lambda_L^\ast s_L^{\nu}  c_R^\ell c^G )]\cdot F_{LR}^{S,3}(\frac{m_{\mu}^2}{m_{\delta^-}^2}) \, .
\end{align}
Further analyses and simplifications of the analytical results have be discussed in subsection \ref{subsection_Analytical_results} and we do not repeat here.

As mentioned in the subsection \ref{subsec_overview_main_physics}, internal scalars in Fig. \ref{fig-muongm2VLL_Feynman} can be replaced by SM gauge bosons when applicable. In Type-II seesaw model, the gauge boson contributions are the same as those in the SM because of the same neutral and charged gauge interactions. In Type-III seesaw model, the gauge boson contributions can deviate from those in the SM because of the modified neutral and charged gauge interactions. In our model, the total new physics contributions from gauge boson Feynman diagrams are negligible compared to the chirally enhanced contributions from scalar mediators.

Based on our calculation we can also reproduce the results of models including only one scalar triplet or one fermion triplet corresponding to Type-II and Type-III seesaw models, respectively. If keeping only the scalar triplet or lepton triplet in this Appendix by turning off relevant interactions, we can reproduce the analytical results of $a_\mu$ predicted in Type-II and Type-III seesaw models.
\begin{itemize}
\item By taking $ s_L^\ell = s_R^\ell =s_L^{\nu}=\lambda_L=0$, the $a_{\mu}^{F,\,\textrm{total}}$ vanishes automatically and we would reproduce the results in Type-II seesaw model as follows, which are negative and consistent with \cite{Freitas:2014pua, Queiroz:2014zfa, Lindner:2016bgg, Athron:2021iuf},
\begin{align}
a_{\mu}^{\ell,\,\textrm{total}}&=\frac{m_{\mu}^2}{8\pi^2}\big\{-\frac{2m_{\mu}^2(s^h)^2}{m_h^2v_h^2}\cdot[F_{LL}^{f,1}(\frac{m_{\mu}^2}{m_h^2})+F_{LR}^{f,1}(\frac{m_{\mu}^2}{m_h^2})]+\frac{2m_{\mu}^2(s^h)^2}{m_{\delta^0}^2v_h^2}\cdot[F_{LL}^{f,1}(\frac{m_{\mu}^2}{m_{\delta^0}^2})+F_{LR}^{f,1}(\frac{m_{\mu}^2}{m_{\delta^0}^2})]\nonumber\\
&\qquad\quad\,\,\,\, +\frac{4|x_L|^2}{m_{\delta^{--}}^2}\cdot[-F_{LL}^{f,1}(\frac{m_{\mu}^2}{m_{\delta^{--}}^2})+2F_{LL}^{S,1}(\frac{m_{\mu}^2}{m_{\delta^{--}}^2})]\big\} \, ,\nonumber\\
&\approx \frac{m_{\mu}^2}{8\pi^2}\big\{-\frac{2m_{\mu}^2(s^h)^2}{m_h^2v_h^2}[\log(\frac{m_h}{m_{\mu}})-\frac{7}{12}]+\frac{2m_{\mu}^2(s^h)^2}{m_{\delta^0}^2v_h^2}[\log(\frac{m_{\delta^0}}{m_{\mu}})-\frac{7}{12}]-\frac{4|x_L|^2}{3m_{\delta^{--}}^2}\big\} \, , \nonumber \\
a_{\mu}^{\nu,\,\textrm{total}}&=\frac{m_{\mu}^2|x_L|^2}{4\pi^2m_{\delta^-}^2}\cdot F_{LL}^{S,3}(\frac{m_{\mu}^2}{m_{\delta^-}^2})\approx-\frac{m_{\mu}^2|x_L|^2}{48\pi^2m_{\delta^-}^2} \, .
\end{align}
\item By taking $s^h=s^a=s^G=x_L=\lambda_L=0$, the $a_{\mu}^{\nu,\,\textrm{total}}$ vanishes automatically and we would reproduce the scalar contributions in Type-III seesaw model as follows. As studied in Refs. \cite{Biggio:2008in, Lindner:2016bgg}, the total contribution in Type-III seesaw model is also negative when combining the scalar and gauge boson contributions.
\begin{align}
a_{\mu}^{\ell,\,\textrm{total}}&= \frac{m_{\mu}^4}{4\pi^2m_h^2v_h^2}[( c_L^\ell )^4-1]\cdot[F_{LL}^{f,1}(\frac{m_{\mu}^2}{m_h^2})+F_{LR}^{f,1}(\frac{m_{\mu}^2}{m_h^2})]\nonumber\\
&\approx  -\frac{m_{\mu}^4( s_L^\ell )^2}{2\pi^2m_h^2v_h^2}(\log(\frac{m_h}{m_{\mu}})-\frac{7}{12}) \, , \nonumber \\
a_{\mu}^{F,\,\textrm{total}}&= \frac{m_{\mu}^2m_{F^{\pm}}^2}{8\pi^2m_h^2v_h^2}( s_L^\ell )^2( c_L^\ell )^2[(1+\frac{m_{\mu}^2}{m_{F^{\pm}}^2})F_{LL}^{f,2}(\frac{m_{F^{\pm}}^2}{m_h^2})+2F_{LR}^{f,2}(\frac{m_{F^{\pm}}^2}{m_h^2})]\nonumber\\
&\approx  \frac{m_{\mu}^2m_{F^{\pm}}^2}{8\pi^2m_h^2v_h^2}( s_L^\ell )^2( c_L^\ell )^2[F_{LL}^{f,2}(\frac{m_{F^{\pm}}^2}{m_h^2})+2F_{LR}^{f,2}(\frac{m_{F^{\pm}}^2}{m_h^2})] \, .
\end{align}
\end{itemize}

\section{Estimations on the corrections to muon mass}\label{sec-estimation-corrections-to-muon-mass}

For the general muon Yukawa interaction of $(y_L\overline{\mu_R} f_L+y_R\overline{\mu_L}f_R)S+\mathrm{h.c.}$, the muon mass correction at one-loop is in the form of 
$m_{\mu}(|y_L|^2+|y_R|^2)$ and $m_f\left[y_L(y_R)^\ast+y_R(y_L)^\ast\right]$.
For $m_{\mu}\ll m_f$, the $m_{\mu}(|y_L|^2+|y_R|^2)$ part is usually small. Then, the loop corrections to the muon mass can be naively taken as 
\begin{align}
\delta m\sim\frac{1}{16\pi^2}m_f\left[y_L(y_R)^\ast+y_R(y_L)^\ast\right],
\end{align}
which agrees with the estimation in Ref. \cite{Czarnecki:2001pv}.\\

We can make a rough estimation for our model. For gauge eigenstate interactions in \eqref{eq-LYukawa-FL}, the relevant interactions are $x_L^{ij}\overline{L_L^i}S\epsilon(L_L^j)^C$ and $\lambda_L^i\overline{\ell_R^i}\mr{Tr}\left[F_L S\right]$. Besides, there should be a mixing between $L_L$ and $F_L$ induced by the $z_L^i\overline{L_L^i}(F_L)^{C}\epsilon H^{\ast}$ interaction. Then, the correction is estimated as
\begin{align}
\frac{\delta m}{m_{\mu}}\sim\frac{1}{8\pi^2}\frac{m_F}{m_{\mu}}x_L\lambda_L\frac{z_Lv_h}{m_F}=\frac{x_L\lambda_L}{8\pi^2}\frac{z_Lv_h}{m_{\mu}}.
\end{align}
In our numerical results, we fix $v_\delta=5\mr{GeV}$. If we take $\{\lambda_L, x_L\}$ as input parameters, $z_L$ is determined. Thus, the correction can be simplified as
\begin{align}
&\frac{\delta m}{m_{\mu}}\sim\frac{x_L\lambda_L}{4\pi^2}\sqrt{ - \sqrt{2} \, x_L \frac{v_\delta \, m_F}{m_{\mu}^2}}\approx-20\lambda_L(-x_L)^{\frac{3}{2}}\sqrt{\frac{m_F}{\mr{TeV}}}.
\end{align}
Then, we find that $\delta m/m_{\mu}\sim-0.5$ for $\lambda_L=1, x_L=-0.06,m_F=3\, \mr{TeV}$. Hence, the mass correction is estimated as $\delta m/m_{\mu}\sim\mc{O}(1)$ for the parameters in our numerical analysis.
\end{appendices}
\end{sloppypar}
\bibliography{ref}
\end{document}